\documentstyle[epsfig,amsfonts,amssymb,12pt]{article}

\makeatletter
\@addtoreset{equation}{section}

\newcommand{\R}{{\Bbb R}}

\newcommand{\Z}{{\Bbb Z}}

\newtheorem{theorem}{Theorem}

\def\t{\Theta}

\def\SU{{\rm SU}}
\def\z{{\bar{z}}}
\def\A{{\cal{A}}}
\def\G{{\mathfrak g}}

\begin{document}

\begin{titlepage}

\title{
\hfill\parbox{4cm}{
{\normalsize LPENSL-TH 03/2001}
}
\\
\vspace{15mm}
The Fuzzy Sphere $\star$-Product and Spin Networks}
\author{
Laurent {\sc Freidel}\thanks{{\tt 
freidel@ens-lyon.fr}}$\hspace{2mm}{}^a$
and 
Kirill {\sc Krasnov}\thanks{{\tt 
krasnov@cosmic.physics.ucsb.edu}}$\hspace{2mm}{}^b$
\\[10pt]
${}^a${\it Laboratoire de Physique}  \\
{\it Ecole Normale Sup\'erieure de Lyon} \\
{\it 46, all\'ee d'Italie, 69364 Lyon Cedex 07, France}\thanks{{\tt UMR
5672  du CNRS}}\\
${}^b${\it Department of Physics, }\\
{\it University of California, Santa Barbara, CA 93106}\\
}

\date{\today}
\maketitle
\thispagestyle{empty}

\begin{abstract}
We analyze the expansion of the fuzzy sphere non-commutative product in powers 
of the non-commutativity parameter. To analyze this expansion  we develop 
a graphical technique that uses spin networks. This technique is potentially 
interesting in its own right as introducing spin networks of Penrose into 
non-commutative geometry. Our analysis leads to a clarification of the link 
between the fuzzy sphere non-commutative product and the usual deformation 
quantization of the sphere in terms of the $\star$-product. 
\end{abstract}

\end{titlepage}

\section{Introduction}
\label{sec:intr}

This paper originated from an observation that manipulating with 
functions on the fuzzy sphere is equivalent to manipulating with
certain $\SU(2)$ spin networks. Although this observation is nothing
more than a reinterpretation of the construction \cite{Hoppe} 
of non-commutative sphere spherical harmonics, it does bring
spin networks of Penrose \cite{Penrose}
into the subject of non-commutative geometry, and is
thus interesting as providing an unusual perspective on non-commutative
manifolds.

In this paper we would like to illustrate the usage of spin networks
by deriving some facts about the non-commutative product on the fuzzy sphere. 
The fuzzy sphere of \cite{Madore} gives one of the simplest 
examples of non-commutative spaces. Its structure is, in a
sense, even simpler than that of the usual non-commutative plane,
for the algebra of functions on the fuzzy sphere is {\it finite
dimensional}, unlike in the case of the plane. On the other hand,
as we shall see, the structure of the $\star$-product in this 
case is much more complicated. 

In \cite{Madore} the fuzzy sphere is  
constructed replacing the algebra 
of polynomials on the sphere by the non-commutative
algebra generated by Pauli matrices taken in a fixed irreducible
representation of ${\rm SU}(2)$. More precisely, the algebra of functions
from $L^2(S^2)$ is thought
of as the algebra of polynomials in $x_i\in\R^3$ modulo the relation
$\|x\|^2 = 1$. We set the radius of the sphere to be $1$. 
One then quantizes the coordinates $x_i$ via:
\begin{equation}
x_i \to \hat{X}^i := {\hat{J}^i \over\sqrt{{N\over2}({N\over2}+1)}} ,
\label{quant}
\end{equation}
where $\hat{J}^i$ are the generators of ${\mathfrak su}(2)$, satisfying
$[\hat{J}^i,\hat{J}^j]=i\epsilon^{ijk}\hat{J}^k$, taken in the
$(N+1)$-dimensional irreducible representation of ${\rm SU}(2)$. 
The factor in (\ref{quant}) is adjusted precisely in such a way
that the ``quantized'' coordinates $\hat{X}^i$ square to one.
The rule (\ref{quant}) gives quantization of the monomials of order
one in coordinates. Monomials of order up to $N$ are quantized by 
replacing the  products of the coordinates $x_i$ by the 
symmetrized products of matrices $\hat{X}^i$ (see more on this map in
Section \ref{sec:star}). Monomials of order
$(N+1)$ and higher become linearly dependent of lower order monomials, 
so that the algebra of functions on the fuzzy sphere is finite-dimensional. 
The integral over $S^2$ is replaced by the trace:
\begin{equation}
\int_{S^2} \to {1\over (N+1)} {\rm Tr}\, .
\label{trace}
\end{equation}
When $1/N$, which plays the role of the parameter of non-commutativity,
is taken to zero, the commutator of $\hat{X}^i$ vanishes, and the
algebra of functions on the non-commutative sphere reduces to the
commutative algebra. The trace (\ref{trace}) then reduces
to the usual integral. It is in this sense that the fuzzy sphere
reduces to the usual $S^2$ when the non-commutativity parameter
is sent to zero. For more details on this construction see, e.g,
\cite{Madore}.

In practice one would like to have a more explicit description of
the above quantization map. In particular, any function on the
sphere can be decomposed into the basis of spherical harmonics, and
one would like to know the matrices into which the spherical
harmonics are sent under the quantization map. This was described
in \cite{Hoppe}, where it was realized that the components of these 
matrices are given by certain Clebsch-Gordan coefficients. Interestingly, 
the work \cite{Hoppe} appeared before the introduction
of the notion of fuzzy sphere in \cite{Madore}, and contained 
essentially all the ingredients of the construction.

Let us now explain why one should expect the appearance of Clebsch-Gordan
coefficients, or $3j$-symbols, as the result of the quantization map.
Consider the following simple chain of isomorphisms:
\begin{equation}
{\rm End}(V^{N\over2}) \sim V^{N\over2}\times (V^{N\over2})^* 
\sim \oplus_{l=1}^N V^l \, .
\label{chain}
\end{equation}
Here $V^l$ is the space of the irreducible representation of ${\rm SU}(2)$ of 
the dimension $(2l+1)$. The Clebsch-Gordan coefficients relate
a basis on the right hand side of (\ref{chain}) to a basis on the
left hand side. A particular basis in $V^l$ is given
by the usual spherical harmonics $Y_{lm}(\theta,\phi)$. Thus, the
isomorphism in (\ref{chain}) implies that every $Y_{lm}, l\leq N$ must be
representable as an element of ${\rm End}(V^{N\over2})$, that is, as
an $(N+1)\times(N+1)$ matrix. The components of this matrix are
given by the corresponding Clebsch-Gordan coefficients. We shall
write down the corresponding formulas in the next section. 

The appearance of the Clebsch-Gordan coefficients, or $3j$-symbols,
as components of the matrices representing the spherical
harmonics indicates that spin networks must play some role.
Indeed, spin networks are exactly the quantities constructed
from $3j$-symbols, corresponding to their vertices, with their
indices contracted in some way, the contraction being represented
by the spin network edges. As we shall see, the problem of calculation of the
non-commutative analog of the integral of a product of a number
of spherical harmonics always reduces to the evaluation of a
particular spin network. One of the goals of the present paper is to develop
the corresponding techniques. We do this by studying in some detail
the $\star$-product on the fuzzy sphere.  

Our paper partially overlaps with work \cite{Alekseev}.
In particular, the formula for the $\star$-product on the
fuzzy sphere in terms of the $6j$-symbol is also contained in 
\cite{Alekseev}. What is new in this paper is the expression
for the expansion of this $\star$-product in powers of the
non-commutativity parameter.
Our approach also allows us to clarify the link between the fuzzy 
sphere product and the deformation quantization $\star$-product.

\section{Setup}
\label{sec:setup}

Under the quantization map the spherical harmonics $Y_{lm}(x), x\in S^2$
are mapped into certain $(N+1)\times(N+1)$ matrices, and, as we
explained in the introduction, the components of these matrices are
given by the Clebsch-Gordan coefficients. Before we spell out what
these matrices are, let us fix our conventions as to what basis
of spherical harmonics is used. Let us introduce, for integer $l$,
\begin{equation}
\bar{\t}^l_m(x) := i^{-m} \langle l,m | T_g | \omega \rangle,
\label{theta}
\end{equation}
where bar denotes the complex conjugation,  
$| l,m \rangle$ form a highest weight normalized basis 
in the irreducible representation
of the dimension $(2l+1)$, and $|\omega\rangle$ is the vector (unique
up to a phase) that is invariant under the action of some fixed ${\rm SO}(2)$
subgroup of ${\rm SO}(3)$. It is given by $|\omega\rangle = 
|l , 0\rangle$. Then (\ref{theta}) is a function on
the homogeneous space ${\rm SO}(3)/{\rm SO}(2)\sim S^2$. Using the
formula (\ref{int-2}) for the integral of the product of two matrix
elements, one gets the orthogonality relation for $\t$:
\begin{equation}
\int_{S^2} dx\,\bar{\t}^l_m \t^{l'}_{m'} = 
{\delta^{ll'} \delta_{mm'}\over {\rm dim}_l},
\label{norm}
\end{equation}
where $dx$ is the normalized measure on the sphere. 
The presence of the factor of ${\rm dim}_l = (2l+1)$ in this formula
(and also the usage of the normalized measure $dx$)
is what makes our $\t^l_m$ different from the usual spherical harmonics
$Y_{lm}$. The basis (\ref{theta}) satisfies:
\begin{equation}
\bar{\t}^l_m = (-1)^m \t^l_{-m}.
\label{conj}
\end{equation}
Note that the same relation is satisfied by the usual $Y_{lm}$, so our basis
is indeed only different by a normalization.

Along with the orthogonality relation (\ref{norm}), 
we will need the value of the
integral of the product of three spherical harmonics. It is easily computed
using the formula (\ref{int-3}) for the integral of the product of three
matrix elements and the definition (\ref{theta}) of the spherical harmonics.
We have:
\begin{equation}
\int_{S^2} dx \, \overline{\t^{l_3}_{m_3}(x)} \t^{l_1}_{m_1}(x)
\t^{l_2}_{m_2}(x) =  
\overline{\hat{C}^{l_3 l_1 l_2}_{0\,0\,0}}
\hat{C}^{l_3 l_1 l_2}_{m_3 m_1 m_2}.
\label{integral}
\end{equation}
Here $\hat{C}$ are Clebsch-Gordan coefficients, and we have
used the fact that the right hand side is only non-zero when
$m_1+m_2=m_3$. Our choice of the normalization for Clebsch-Gordan's
is such that the so-called theta-symbol is
always equal to one, see (\ref{theta-1}). For reference, let us
mention that our coefficients $\hat{C}^l$ are given 
by $1/\sqrt{{\rm dim}_l}$ times the Clebsch-Gordan 
coefficients used by Vilenkin and Klimyk \cite{VK}. The ``hat''
over the symbol of the coefficient is used precisely to 
indicate this difference in normalizations.
It is now not hard to see that (\ref{integral}), together
with (\ref{norm}), implies that:
\begin{equation}\label{sign}
\t^{l_1}_{m_1}(x) \t^{l_2}_{m_2}(x) = \sum_{l_3=|l_1-l_2|}^{l_1+l_2} \sum_{m_3} {\rm dim}_{l_3}
\hat{\t}^{l_3}_{m_3}(x) \overline{\hat{C}^{l_3 l_1 l_2}_{0\,0\,0}} 
\hat{C}^{l_3 l_1 l_2}_{m_3 m_1 m_2}.
\label{class-prod}
\end{equation}
Note that, in this formula, the commutativity of the product comes from 
the symmetry relations for the Clebsch-Gordan, namely:
\begin{equation}
\hat{C}^{l_3 l_2 l_1}_{m_3 m_2 m_1}=
(-1)^{l_3 -l_1 -l_2} \hat{C}^{l_3 l_1 l_2}_{m_3 m_1 m_2}.
\end{equation}

The group ${\rm SO}(3)$ acts on functions on $S^2$ by left shifts 
$T_g f(x) = f(g^{-1} x)$, and the functions $\t^l_m(x)$ for fixed
$l$ span the vector space $V^l$, in which the representation
by shifts is irreducible. In view of the isomorphisms (\ref{chain}),
functions $\t^l_m(x), l\leq N$ can be mapped to $(N+1)\times(N+1)$
matrices, whose components must be given by Clebsch-Gordan coefficients.
These matrices are:
\begin{equation}
[\hat{\t}^l_m]_{ij} = \sqrt{N+1} \,\, \hat{C}^{l {N\over 2} {N\over 2}^*}_{m\,i\, j},
\label{theta-q}
\end{equation} 
where $\hat{C}^{l {N\over 2} {N\over 2}^*}_{m\, i\, j}$ are Clebsch-Gordan coefficients
with properties:
\begin{eqnarray} \label{norm-q}
\sum_{mij} \overline{\hat{C}^{l {N\over 2} {N\over 2}^*}_{m\, i\, j}}
\hat{C}^{l {N\over 2} {N\over 2}^*}_{m\, i\, j} = 1 \\
\overline{\hat{C}^{l {N\over 2} {N\over 2}^*}_{m\, j\, i}} = 
(-1)^m \hat{C}^{l {N\over 2} {N\over 2}^*}_{-m\, i j} .
\end{eqnarray}
Here $(N/2)^*$ is the conjugate representation to $N/2$.
The first of these properties is the normalization condition that
will be explained below, while the second is the ``quantum''
analog:
\begin{equation}
(\hat{\t}^l_m)^\dagger = \widehat{\overline{\t^l_m}}
\end{equation}
of the classical property (\ref{conj}). 

We are now in the position to introduce the graphic notations that
will lead us to spin networks and somewhat explain the normalization
choices made above. We shall denote the Clebsch-Gordan's by
a tri-valent vertex, with its three edges representing the three pairs of
indices of $\hat{C}$, so that
\begin{equation}
[\hat{\t}^l_m]_{ij} = \sqrt{N+1} 
\qquad\lower0.3in\hbox{\epsfig{figure=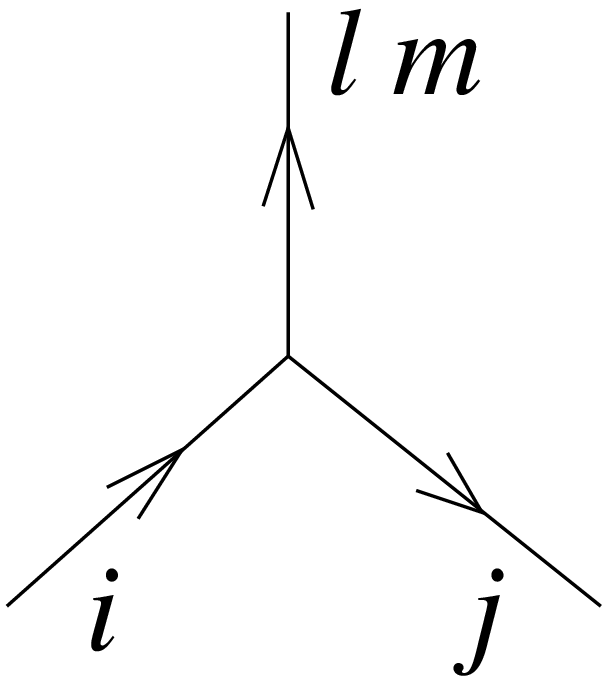,height=0.6in}}
\label{3j}
\end{equation}
Each edge corresponds to a pair of indices: an irreducible representation
(spin) and a basis vector in this representation. If no spin is
indicated, as for the bottom edges in (\ref{3j}), then $N/2$ is
assumed. Edges are oriented. The operation of complex conjugation
is represented by changing the orientation of all the edges. The 
origin of the graphic notation (\ref{3j}) is evident: the Clebsch-Gordan
coefficients it represents are just the matrix elements of the intertwiner
between the tensor product of $V^{N\over2}\times (V^{N\over2})^*$ and the
representation $V^l$; this intertwiner is represented by a trivalent vertex.

Using the graphical notation introduced, the normalization condition
(\ref{norm-q}) is the statement about the value of the so-called
theta graph:
\begin{equation}\label{theta-1}
\lower0.2in\hbox{\epsfig{figure=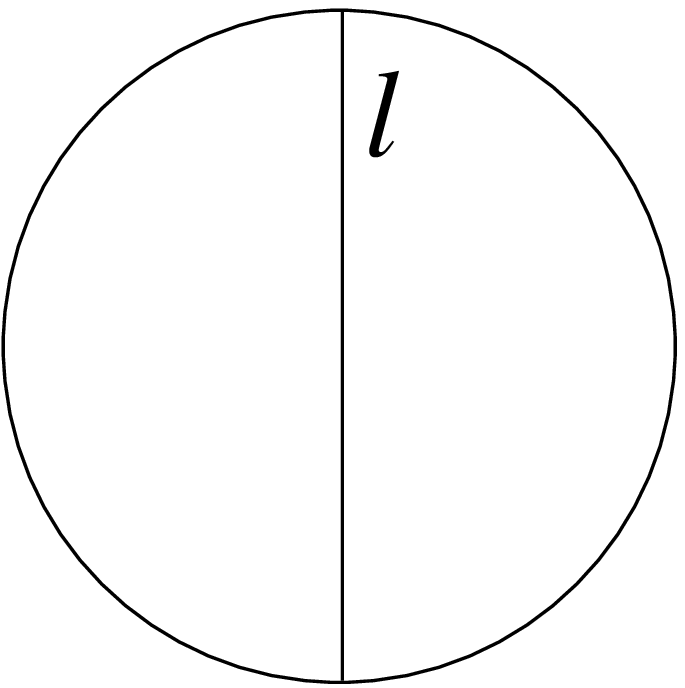,height=0.5in}} = 1.
\end{equation}
This graph is constructed taking the product of two $3j$-symbols
and summing over the ``internal'' indices, as in (\ref{norm-q}).
The theta graph is the simplest spin network, and the $3j$
symbols we use are normalized in precisely such a way that
the value of this graph is always one. 

As we mentioned in the Introduction, for the non-commutative sphere,
the integral over $S^2$ goes into the trace. Let us now find the
quantum analog of the orthogonality relation (\ref{norm}). The
corresponding quantity is graphically represented as:
\begin{equation}
{1\over N+1} {\rm Tr} \left( \hat{\t}^l_m \right)^\dagger \hat{\t}^{l'}_{m'} 
= \lower0.4in\hbox{\epsfig{figure=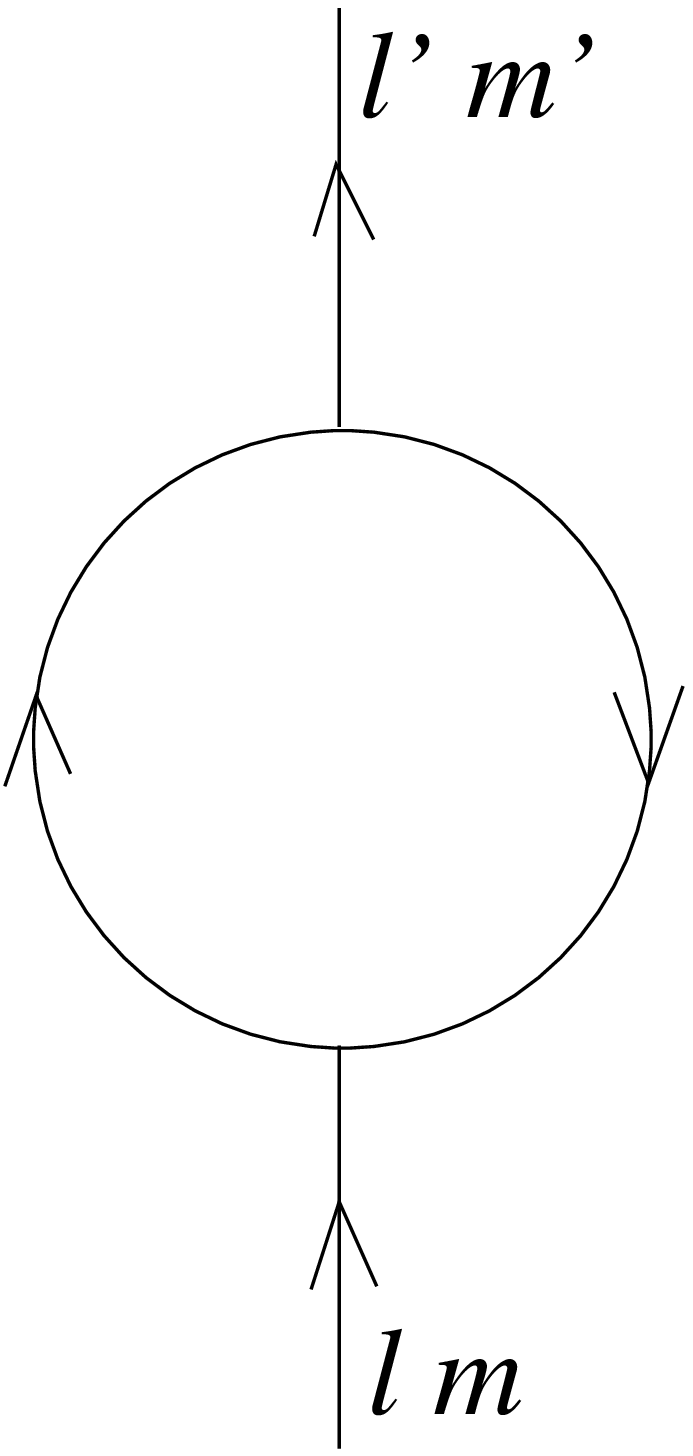,height=0.9in}}
\end{equation}
Here $\dagger$ denotes the operation of taking the Hermitian conjugation,
and its effect is exactly such that all ``internal'' indices
are contracted as in the above diagram. Now using the elementary
fact that:
\begin{equation}
\lower0.4in\hbox{\epsfig{figure=ort.eps,height=0.9in}} =
{1\over {\rm dim}_l} \,\,\, 
\lower0.3in\hbox{\epsfig{figure=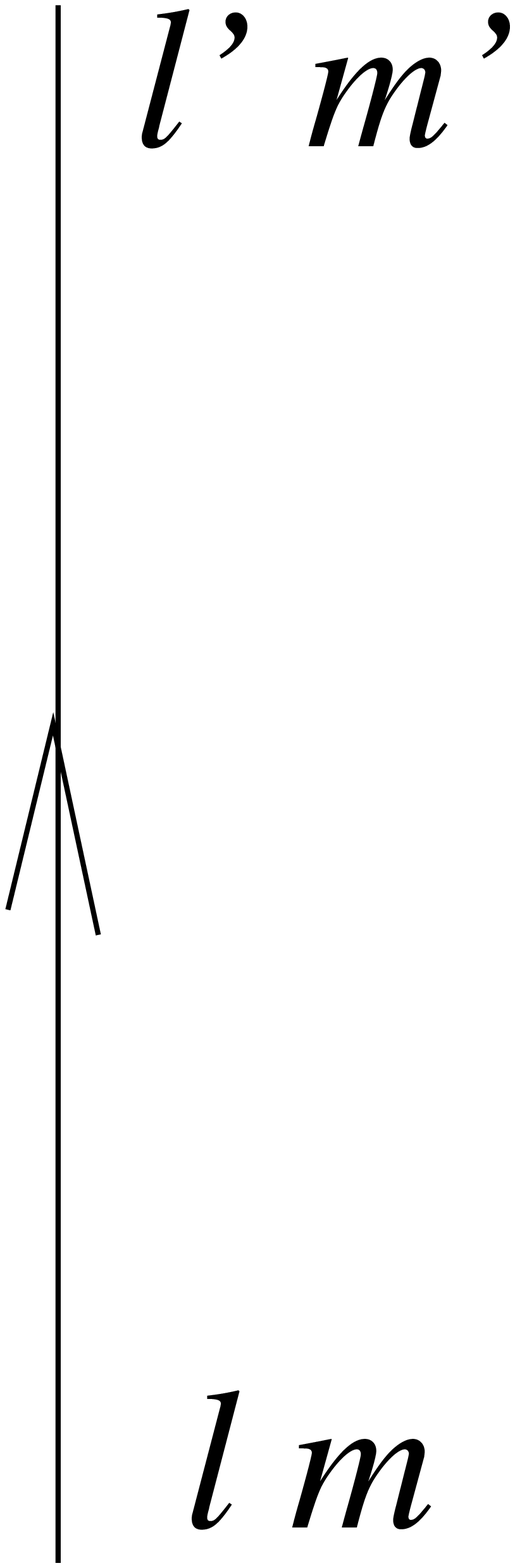,height=0.72in}}
\,\,\, \lower0.2in\hbox{\epsfig{figure=theta.eps,height=0.5in}}
\label{tt}
\end{equation}
where the straight line represents the matrix element of the intertwiner
between representations $l$ and $l'$, that is, the product of
Kronecker deltas, we see that the ``quantum'' spherical harmonics
$\hat{\t}$ satisfy exactly the same orthogonality relation (\ref{norm})
as the classical ones. The factor of $\sqrt{N+1}$ in (\ref{theta-q})
was adjusted precisely in such a way that this property holds.

We hope that the reader has already started appreciating the convenience
of the graphical notations. Using these notations, complicated 
expressions are calculated by using elementary facts from
representation theory. Thus, the property (\ref{tt}) is proved
by noticing that the left hand side of this equation is an intertwiner
between representations $l$ and $l'$. Such an intertwiner only
exists when $l=l'$, which means that the result must be proportional
to the straight line. The proportionality coefficient can be calculated
by taking the trace of the whole expression, which is graphically represented
by ``closing'' the open ends of the diagram. Performing this operation
on the left hand side, one gets the theta graph. The right hand side gives
the loop, whose value is just the dimension of the corresponding 
representation. We shall see other examples of such proofs in the
sequel.

Let us now, before we go to our discussion of the $\star$-product on
the fuzzy sphere, prove that the quantization rule (\ref{theta-q}) does
give the correct quantization of the sphere, that is, the one
given by (\ref{quant}). To this end, we must calculate the commutator
of the non-commuting coordinates $\hat{X}^i$. Recall that classically
the coordinates $x_i$ are just the spherical harmonics corresponding
to $l=1$: 
\begin{eqnarray}
x_1 = {1\over\sqrt{2}}(\t^1_1 - \t^1_{-1}), \qquad
x_2 = {1\over i \sqrt{2}}(\t^1_1 + \t^1_{-1}), \qquad
x_3 = \t^1_0.
\end{eqnarray}
In the quantum case we replace the harmonics $\t^1$ by the
corresponding matrices (\ref{theta-q}). We then must have:
\begin{equation}
[\hat{X}^{i},\hat{X}^{j}] = i \epsilon^{ijk} \hat{X}^{k} 
{1\over\sqrt{{N\over2}({N\over2}+1)}}.
\label{comm-t}
\end{equation}
Let us show that this property indeed holds. We have:
\begin{equation}\label{comm'}
{1\over N+1} [\hat{X}^{i},\hat{X}^{j}] = 
\lower0.2in\hbox{\epsfig{figure=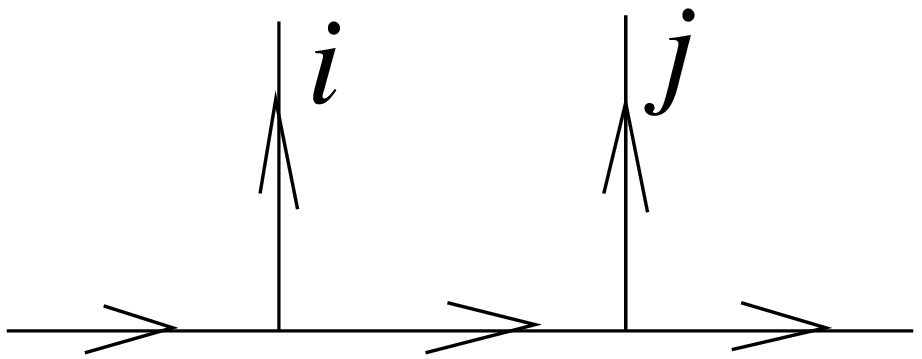,height=0.5in}} -
\lower0.2in\hbox{\epsfig{figure=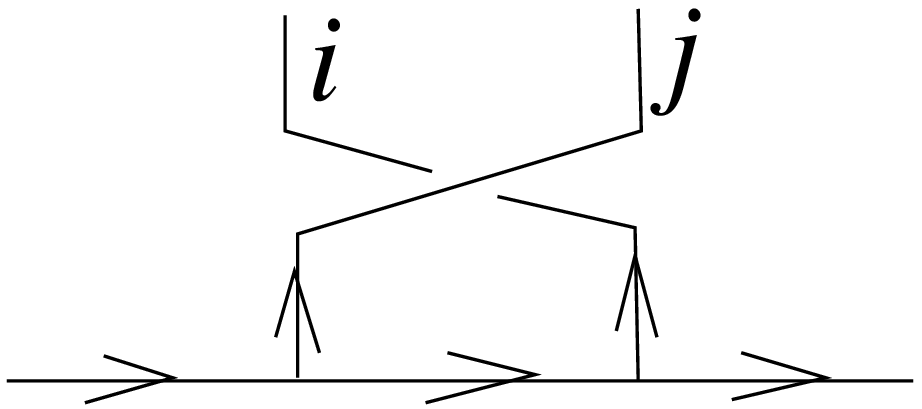,height=0.5in}} 
\end{equation}
where the spin $1$ is assumed on the vertical lines, and
\begin{equation} \label{comm}
{1\over (N+1)^{3/2}} {\rm Tr} \left( (\hat{X}^{k})^\dagger
[\hat{X}^{i},\hat{X}^{j}] \right)  = 
\lower0.35in\hbox{\epsfig{figure=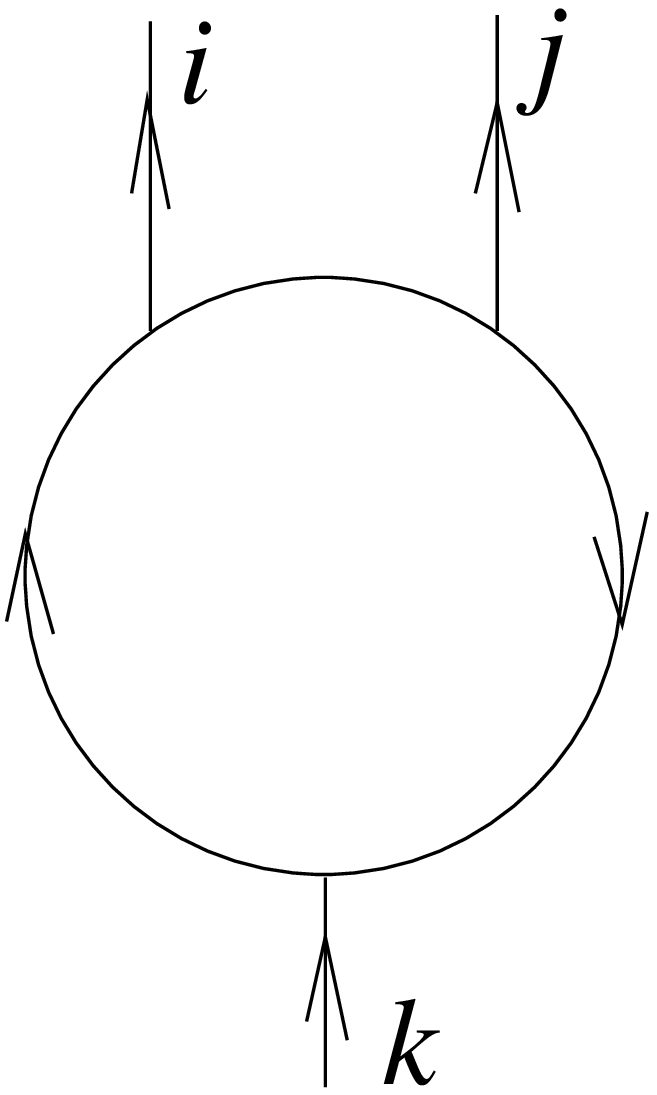,height=0.7in}}
- \lower0.35in\hbox{\epsfig{figure=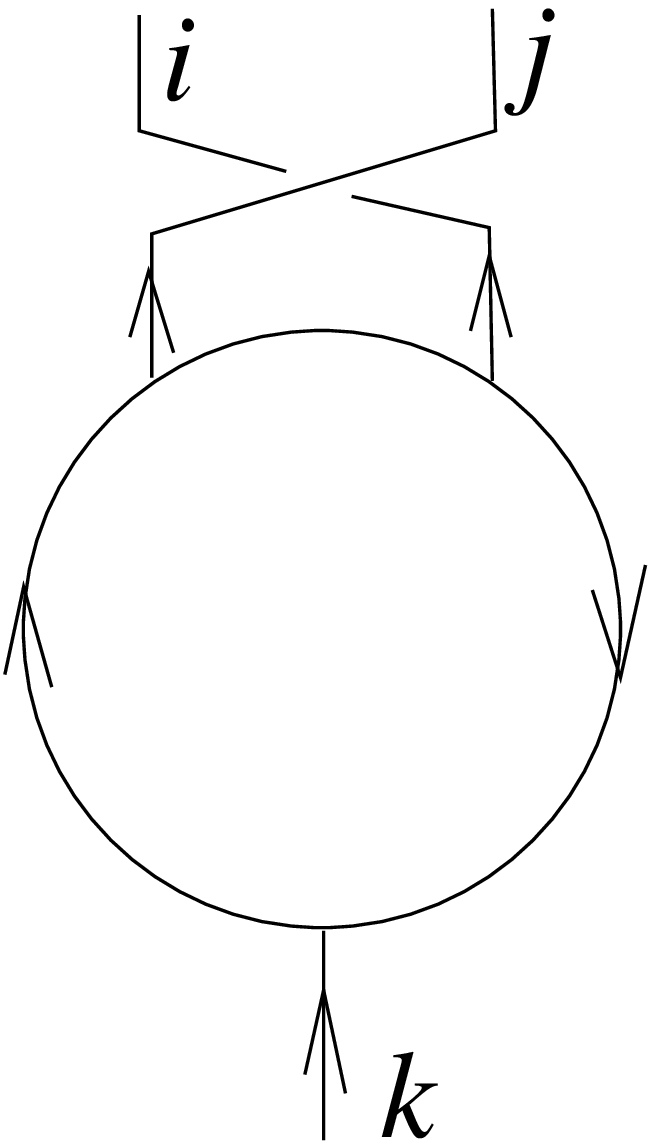,height=0.7in}} = 
2 \lower0.3in\hbox{\epsfig{figure=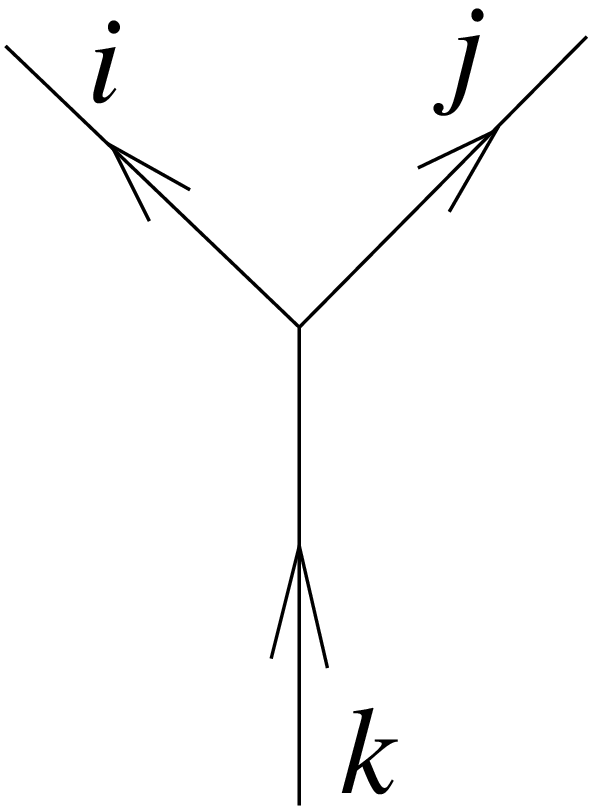,height=0.6in}}
\lower0.2in\hbox{\epsfig{figure=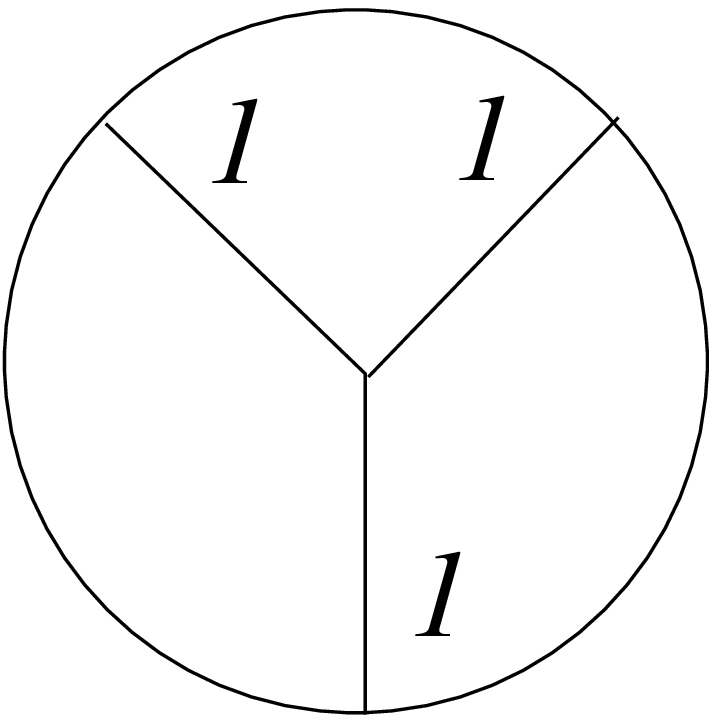,height=0.5in}}
\end{equation}
To get the last equality we used the fact that each of the two 
terms is an intertwiner from a single representation of spin 1
to the tensor product of two such representations. Such an
intertwiner must be proportional to the unique intertwiner that
is given by the tri-valent vertex on the right hand side of (\ref{comm}). 
The proportionality coefficient is easily determined by closing
all the open edges, and is given by the $6j$-symbol on the 
right hand side. The second term equals minus the first one see 
(\ref{sign}), which
explains the factor of $2$. The $3j$-symbol is given by:
\begin{equation}
\lower0.3in\hbox{\epsfig{figure=tr-comm2-1.eps,height=0.6in}} =
{(-i)\over \sqrt{3!}} \epsilon^{ijk},
\end{equation}
and the $6j$-symbol can be calculated using the formula (\ref{app:6j})
given in the Appendix. The result is:
\begin{equation}\label{111}
\lower0.2in\hbox{\epsfig{figure=tr-comm2-2.eps,height=0.5in}} =
(-1){1\over\sqrt{3!}} {1\over\sqrt{N+1}} {1\over\sqrt{{N\over2}({N\over2}+1)}}.
\end{equation}
Combining all this together, we see that (\ref{comm-t}) indeed
holds. 

We thus learn that the quantization rule (\ref{theta-q}) coincides
with the standard quantization map (\ref{quant}), at least for the 
spherical harmonics of the first order. Interestingly, the result
of the commutator in the quantum case turns out to be expressed 
through the $6j$-symbol. In the case considered when all the spins
were taken to be $l=1$, the commutator coincides with the classical
result. However, for higher modes $l$ one expects a deviation from
the classical expressions. This is summarized in the notion of the
$\star$-product. We shall now illustrate the described spin network
techniques by deriving some facts about this $\star$-product.

\section{The $\star_N$-product}
\label{sec:star}

As we have said in the introduction, our aim is 
to illustrate the spin network construction by deriving some facts about the 
fuzzy sphere non-commutative product. However, before we introduce
and study this product, let us review the usual $\star$-product that
arises in the deformation quantization of $\R^3$ equipped with
Poisson structure invariant under the rotation group. We will then
discuss a relation between this, and the fuzzy sphere product.

Let us start by reminding some general facts about $\star$-products.
A $\star$-product gives a non-commutative deformation of the usual 
(point-wise) multiplication of functions. 
Let  $\A= {\cal C}^{\infty}(M)$ be the space of smooth function on a manifold 
$M$, and suppose that $M$ is equipped with a Poisson bracket 
$\{\cdot, \cdot\}$.
Let us denote by $A[[\hbar]]$ the space of formal power series with 
coefficients in $\A$. A star product on $\A$ is an associative, 
$\R[[\hbar]]$-linear product on $A[[\hbar]]$. The product of two 
functions on $M$ is given by
\begin{equation}
\phi\star \psi = \sum_{n}\hbar^{n} B_{n}(f,g),
\end{equation}
$ B_{n}(\cdot,\cdot)$ being bidifferential operators.
The first term in the expansion is the usual commutative product 
 $B_{0}(f,g) =fg$, while the first term in the commutator is the 
 Poisson bracket $B_{1}(f,g) -B_{1}(g,f) =\{f,g\}$.
 The Poisson bracket therefore gives the germ of deformation of the 
 commutative product towards $\star$-product.
There is a notion of gauge transformation that can be defined on 
the set of $\star$-products and Poisson brackets. 
The group of these gauge transformations consists of linear automorphisms 
of $\A[[\hbar]]$ of the following form:
\begin{equation}
 f\rightarrow U(f)= f + \hbar U_{1}(f)+\cdots +\hbar^{n} U_{n}(f),
\end{equation}
where $U_{i}(f)$ are differential operators.
It acts on the set of star product as
\begin{equation}
    \phi \star_{U} \psi = U^{-1}(U(\phi)\star U(\psi)).
\end{equation}
Products related by a such transformation are called equivalent.

As is well known, if $M$ is the dual of a Lie algebra, $M=\G^{*}$
there is the so-called Kirillov-Lie Poisson structure on it, 
whose symplectic leaves are coadjoint orbits. The bracket is
\begin{equation}
\{\phi, \psi\}(x) = x^{k}C^{ij}_{k} \partial_{i}\phi(x) 
\partial_{j}\psi(x),
\end{equation}
where $x^{i}$ are the components of $x \in \G^{*}$. 
Using the natural identification between linear functions on $\G^{*}$ 
and $\G$ one can equivalently  view $x^{i}$ as  a basis $e^{i}$ in $\G$. 
Then $C^{ij}_k$ are the structure constants with respect to this basis.
This identification extends to a natural isomorphism 
between the space ${\rm  Pol}({\mathfrak g}^*)$ of polynomials in $x^i$  
and symmetric algebra ${\rm  Sym}(\G^{*})$.
There are several equivalent ways to introduce a deformation 
quantization of the Kirillov bracket. We shall refer to the 
arising $\star$-product as Kirillov $\star$-product.
 
\bigskip
\noindent{\bf Universal enveloping algebra}

The standard way to introduce a $\star$-product in $\G^{*}$ is by using
the universal enveloping algebra ${\mathfrak U}(\G)$. Recall
that the universal enveloping algebra ${\mathfrak U}(\G)$
of the Lie algebra $\mathfrak g$ with generators $e^i$ satisfying
\begin{equation}\label{app-1}
[ e^i, e^j ] = \hbar C^{ij}_k e^k,
\end{equation}
where $C^{ij}_k$ are the structure constants, is the algebra generated by all
polynomials in the generators $e^i$ modulo the relation
$XY-YX = [X,Y]$ for all $X,Y\in\mathfrak g$. 
The Poincare-Birkhoff-Witt theorem states
that $\mathfrak U$ is isomorphic to the algebra ${\rm Sym}(\G)\sim 
{\rm  Pol}({\G}^*)$
generated by all completely symmetrized polynomials in the generators:
${\mathfrak U}(\mathfrak g) \sim {\rm Sym}(\mathfrak g)$. Since the
algebra ${\rm Sym}(\mathfrak g)$ is naturally isomorphic to the algebra
${\rm  Pol}({\mathfrak g}^*)$ of polynomial functions on the dual space 
${\mathfrak g}^*$, we have the one-to-one map 
\begin{eqnarray}\nonumber
\sigma: {\rm Pol}({\mathfrak g}^*) \to {\mathfrak U}(\mathfrak g),
\\ \nonumber
x^{k_1} \ldots x^{k_n} \to {1\over n!} \sum_{s} e^{k_{s(1)}} 
\ldots e^{k_{s(n)}},
\end{eqnarray}
where the sum is taken over all permutations $s$ of $n$ integers,
 $x^i$ are linear functionals on $\mathfrak g^{*}$, and $e^{i}$ are the 
corresponding elements in $\G$.
Using the map $\sigma$, one can define the following non-commutative
product on ${\rm Pol}({\mathfrak g}^*)$:
\begin{equation}\label{prod-1}
(\phi\star\psi)(x) = \sigma^{-1} \left [
\sigma(\phi) \sigma(\psi) \right ]. 
\end{equation}
Note that this $\star$-product explicitly depends on the deformation
parameter $\hbar$ because the later enters the commutator (\ref{app-1}).
It is not so straightforward to see that this product can be expanded 
in terms of differential operators. For example, one can prove the
following formula:
\begin{equation}
(e^X \star Y)(x) = e^{X(x)} \left [ {{\rm ad} X\over 1 - e^{-{\rm ad} X}} 
\cdot Y \right](x).
\end{equation}

\bigskip
\noindent{\bf Baker-Campbell-Haussdorf formula}

The Baker-Campbell-Haussdorf formula states that, given two
Lie algebra elements $X, Y$, the product of the corresponding
group elements can be expressed as:
\begin{equation}
e^X \cdot e^Y = e^{X+Y+\Psi(X,Y)},
\end{equation}
where $X, Y, \Psi(X,Y)\in \G$ and 
\begin{equation}
X+Y+\Psi(X,Y) = \sum_{m=1}^\infty {(-1)^{(m-1)}\over m}
\sum_{\vbox{\hbox{$k_i+l_i\geq 1$}\hbox{$k_i\geq 0, l_i\geq 0$}}}
{[X^{k_1} Y^{l_1} \cdots X^{k_m} Y^{l_m}]\over k_1! l_1! \cdots k_m! l_m!} ,
\end{equation}
and $[X_{1}\cdots X_{n}]$, $X_{i}\in \G $ denotes 
${1\over n} [\cdots[[X_1,X_2],X_3],\cdots,X_n]$. This is the so-called
Campbell-Hausdorff formula in the Dynkin form, see, e.g., \cite{Kirillov}.
Representing the commutator as $[X,Y]=XY-YX$, one can think of
$\Psi(X,Y)$ as a complicated sum of polynomials in the components
$X_i, Y_i$:
\begin{equation}
\Psi_i (X,Y) = \sum_{i_{\vec{n}}\, j_{\vec{m}}}
A_i^{i_{\vec{n}}\, j_{\vec{m}}} X_{i_1} \cdots X_{i_n}
Y_{j_1} \cdots Y_{j_m}.
\end{equation}
One then defines a $\star$-product on the space of functions
on $\R^3={\mathfrak g}^*$ by:
\begin{equation}\label{prod-2}
(\phi\star\psi)(x) = \phi \, e^{{1\over\hbar} 
(x|\Psi(\hbar\partial,\hbar\partial))} \psi,
\end{equation}
where we have replaced the argument $X_i$ of $\Psi(X,Y)$ by
$\hbar$ times the derivative $\partial/\partial x^i$ acting on
the left on $\phi$, and the argument $Y_i$ by the derivative
acting on the right on $\psi$. The quantity $(x|\Psi)$ in the
exponential stands for the contraction $x^i \Psi_i$. One can prove
that the product (\ref{prod-2}) is, in fact, the same as
(\ref{prod-1}). A good exposition of the relation between these
two quantizations, and also of the relation of both to the
Kontsevich deformation quantization, is given in \cite{Vinay}.

\bigskip
\noindent{\bf Path integral}

The third way to define Kirillov $\star$-product uses a version of 
Cattaneo-Felder \cite{CF} path integral. For the case in question,
the function $\alpha^{ij}(x)$ that determines the Poisson structure:
\begin{equation}
\{ \phi, \psi \} = \sum_{ij} \alpha^{ij}(x) \partial_i \phi(x)
\partial_j \psi(x)
\end{equation}
is linear in $x$, and the action used in \cite{CF} can be given a
simple form of the action of the so-called BF theory. Thus, let us
consider the following theory on the unit disc. It has two dynamical 
fields: the field $B_i$, which is ${\mathfrak g}^*$ valued, this 
field is the analog of the field $X^i$ of \cite{CF}, and the $G$-connection 
$A^i$, with curvature $F^i(A)$. This connection is the analog
of the one-form field $\eta$ of \cite{CF}. The action is then given by:
\begin{equation}\label{BF}
S[B,A] = \int_U B_i F^i(A),
\end{equation}
where the integral is taken over the unit disc $U$. The $\star$-product is
then given by the following path integral:
\begin{equation}\label{prod-3}
(\phi\star\psi)(x) = \langle \phi(B(0)) \psi(B(1)) \rangle_x \equiv
\int_{B(\infty)=x} {\cal D}B {\cal D}A \, \phi(B(0)) \psi(B(1))\,
e^{{1\over\hbar} \int BF}.
\end{equation}
Thus, the $\star$-product is obtained by computing the correlation
function of two operators in this theory. The operators are given
by functions $\phi,\psi$ on the values of the $B$-field at two 
boundary points $0,1\in\partial U$. The value of $B$ at the third
boundary point $\infty$ is kept fixed in the path integral. One
also has the boundary conditions for the connection: it is required
that the connection one-form vanishes on $\partial U$ on vectors
tangent to the boundary. Then the perturbative expansion of the
path integral (\ref{prod-3}) gives a non-commutative product
in $\R^3$, which can be checked to be associative, and, thus,
is a $\star$-product. Since the BF action (\ref{BF}) is essentially 
the one used by Cattaneo and Felder \cite{CF}, the product
(\ref{prod-3}) is the Kontsevich product \cite{Kon}. It can then
be shown, see \cite{Vinay}, that this product is equivalent to the one
defined using the Campbell-Hausdorff formula.

\bigskip
\noindent{\bf $\star_N$-Product}

Having reviewed the usual $\star$-product, let us introduce the
non-commutative product that is relevant in the context of the fuzzy sphere. 
The square integrable functions on $S^2$ can be decomposed into 
the basis of spherical harmonics:
\begin{equation}
\phi(x) = \sum_{lm} \phi^l_m \t^l_m(x).
\label{dec}
\end{equation}
We have a quantization map, which sends spherical harmonics
$\t^l_m, l\leq N$ to matrices (\ref{3j}). Under this map,
functions are sent to matrices:
\begin{equation}
\phi(x) \to \hat{\phi}= \sum_{l=0}^N \sum_m \phi^l_m
\hat{\t}^l_m
\label{map}
\end{equation}
Note that this map is insensitive to the ``high frequency'' behavior of
the function, for it cuts off all the harmonics with $l > N$. 

Let us also construct the inverse map. To this end, we introduce
the non-commutative analog of $\delta$-function:
\begin{equation}
\hat{\delta}_x = \sum_{l=0}^N  \sum_m {\rm dim}_l \,
(\hat{\t}^l_m)^\dagger \t^l_m(x).
\label{delta-q}
\end{equation}
Thus, the ``quantum'' $\delta$-function is a matrix that in addition
depends on a point $x\in S^2$. Note that the $\delta$-function
is ``real'': $\hat{\delta}^\dagger = \hat{\delta}$.
Given an arbitrary operator (matrix)
$\hat{\phi}$ one can construct from it a square integrable function:
\begin{equation}
\hat{\phi} \to \phi(x) = {1\over N+1} {\rm Tr} 
\left( \hat{\delta}_x \hat{\phi} \right) .
\label{inverse}
\end{equation}
The resulting functions, of course, contain only the modes with $l\leq N$.
Using the $\delta$-function (\ref{delta-q}) one can also give another
expression for the quantization rule (\ref{map}). Indeed, we have:
\begin{equation}
\hat{\phi} = \int_{S^2} dx\, \hat{\delta}_x \phi(x) .
\end{equation}
One can easily check that the composition of the quantization map
(\ref{map}) and its inverse (\ref{inverse}) give back the function
one started from with all its modes $l>N$ cut off. Let us formalize
this introducing the notion of the projector ${\cal P}_N$:
\begin{equation}
\phi \to \hat{\phi} \to {1\over N+1} {\rm Tr} 
\left( \hat{\delta}_x \hat{\phi} \right) = {\cal P}_N \phi.
\label{proj}
\end{equation}
Thus, the maps (\ref{map}) and (\ref{inverse}) are
one-to-one on the space of functions that only contain modes up to $N$.
Let us denote this space by ${\cal A}^N$. Note that
${\cal A}^N = {\cal A}^\infty/{\rm Ker}\,{\cal P}_N$, 
where ${\cal A}^\infty$ is the algebra of all $L^2$ functions
on $S^2$.

Having defined the quantization map, we can use it to define a non-commutative
product on the space ${\cal A}^N$ via:
\begin{equation}
\widehat{\phi \star_N \psi} = \hat{\phi} \hat{\psi}.
\label{prod-N}
\end{equation}
Let us emphasize that this is well defined only in ${\cal A}^N$ since 
$\hat{\phi}= \widehat{{\cal P}_N \phi}$ and 
 the product $\star_N$ does not coincide with the
usual $\star$-product reviewed above, because 
they are defined on different spaces. While $\star_N$ is the
product on the space of functions ${\cal A}^N$, the ``real'' $\star$-product
acts on the space of all square integrable functions ${\cal A}^\infty$.
However, we are going to show that  the $\star_N$-product is related to 
the usual $\star$-product via:
\begin{equation}
\phi\star_N \psi = {\cal P}_N \left[ \phi{\star_\hbar}\psi\right]
\Big |_{\hbar=2/(N+1)}.
\label{star-1}
\end{equation}
Let us note that the interplay between the non-commutative 
product and deformation quantization was studied before, see, e.g., \cite{Eli}.
What is new in this paper is the justification for the  formula 
(\ref{star-1}) coming from an explicit expression for the  asymptotic expansion of the 
$\star_N$-product in powers of the non-commutativity parameter. 
To find this expansion we consider the $\star_N$-product of the modes. 
The product of two matrices $\hat{\t}$ can be decomposed
into a sum of $\hat{\t}$. A simple calculation, similar to the one performed
in (\ref{comm'}), (\ref{comm}), gives:
\begin{equation}
\t^{l_1}_{m_1} \star_N \t^{l_2}_{m_2} = 
\sum_{l_3=0}^N \sum_{m_3} {\rm dim}_{l_3} \t^{l_3}_{m_3} 
\lower0.2in\hbox{\epsfig{figure=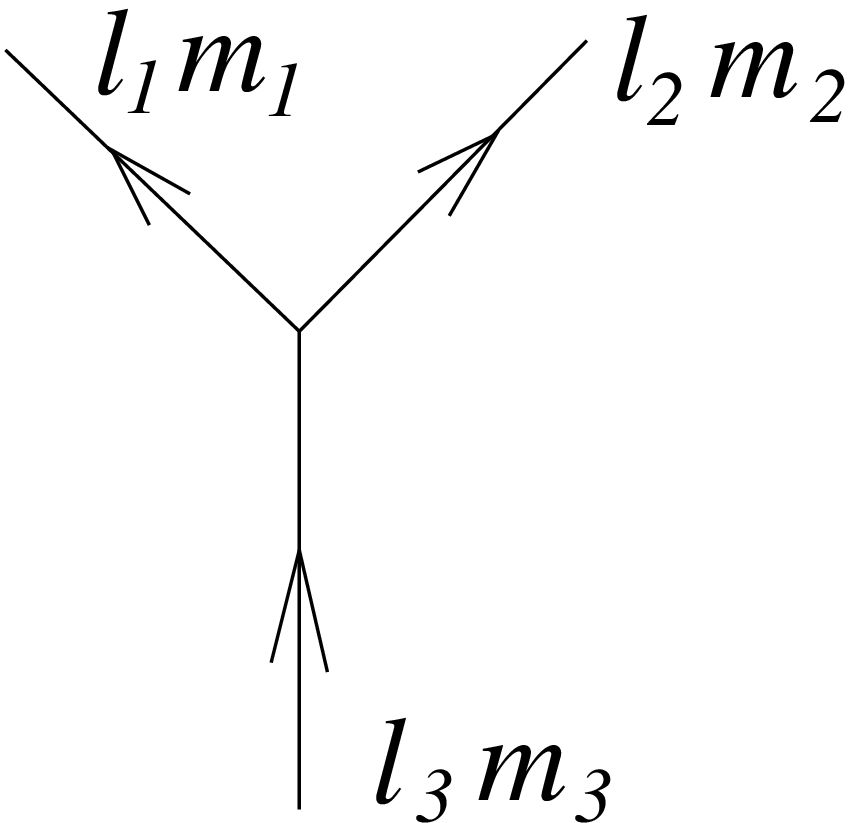,height=0.5in}}
\sqrt{N+1}\,\,\,
\lower0.2in\hbox{\epsfig{figure=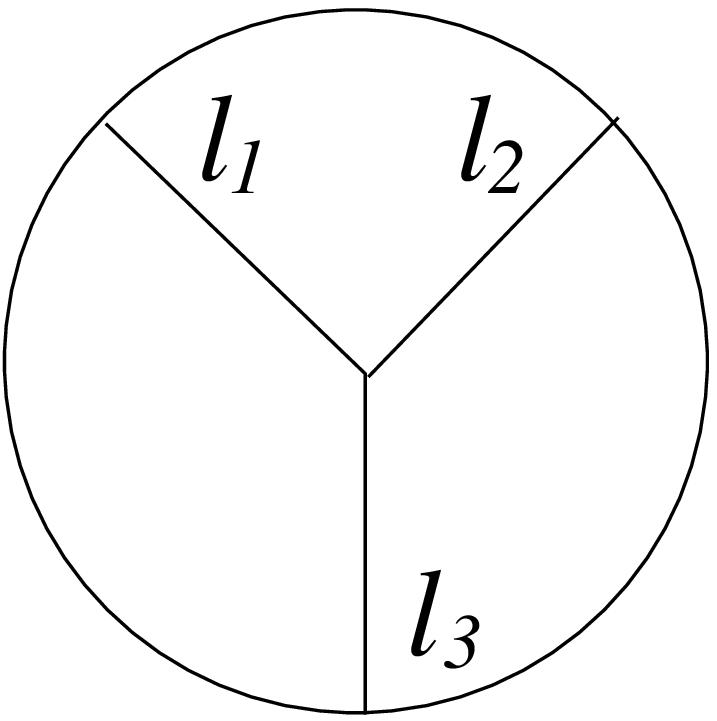,height=0.5in}}
\label{6j}
\end{equation}
Let us notice that the right hand side depends on $N$ only through
the $6j$-symbol (and the square root), and the cutoff in the sum. 
Therefore, the non-trivial information about the
$1/N$ expansion of the $\star_N$-product is all contained in the $6j$-symbol.
Thus, the fuzzy sphere gives us an interesting example of a non-commutative 
manifold for which the $\star$-product is not only known in terms of its $1/N$ 
derivative expansion, but also in closed form, in terms of the
$6j$-symbol in (\ref{6j}). 

Let us note that
(\ref{6j}) can also be written as:
\begin{equation} \label{expr}
\t^{l_1}_{m_1} \star_N \t^{l_2}_{m_2} = 
\sum_{l_3=0}^N {\cal P}^{(l_3)} \left(
\t^{l_1}_{m_1} \cdot \t^{l_2}_{m_2} \right)
\Psi_{(N)}(l_1,l_2,l_3),
\end{equation}
where ${\cal P}^{(l)}$ is the projector on $l$-th representation
\begin{equation}
{\cal P}_N = \sum_{l\leq N} {\cal P}^{(l)},
\end{equation}
and 
\begin{equation}\label{Psi}
\Psi_{(N)}(l_1,l_2,l_3) :=
{\sqrt{N+1}\over \hat{C}^{l_3 l_1 l_2}_{0\,0\,0}} \,\,
\lower0.2in\hbox{\epsfig{figure=lll-2.eps,height=0.5in}}
\end{equation}
As we show below,
\begin{equation}\label{asympt}
\lim_{N\to\infty} \sqrt{N+1} \,\,
\lower0.2in\hbox{\epsfig{figure=lll-2.eps,height=0.5in}}
= \hat{C}^{l_3 l_1 l_2}_{0\,0\,0}.
\end{equation}
This implies that the expansion of
the function $\Psi_{(N)}(l_1,l_2,l_3)$ in powers of $1/N$ starts from one.
This fact, together with the relation (\ref{class-prod}) means
that the zeroth order term in the $1/N$ expansion of the
$\star_N$-product of two modes $\t$ is given by the usual product, which
is what one expects.

Before we look more closely into the details of the asymptotic 
expansion, let us
show that the non-commutative product defined by (\ref{6j})
is indeed an associative product.
Interestingly, the associativity follows from the so-called
Biedenharn-Elliott (or pentagon) identity, which reads:
\begin{equation}
\sum_l {\rm dim}_l \,
\lower0.25in\hbox{\epsfig{figure=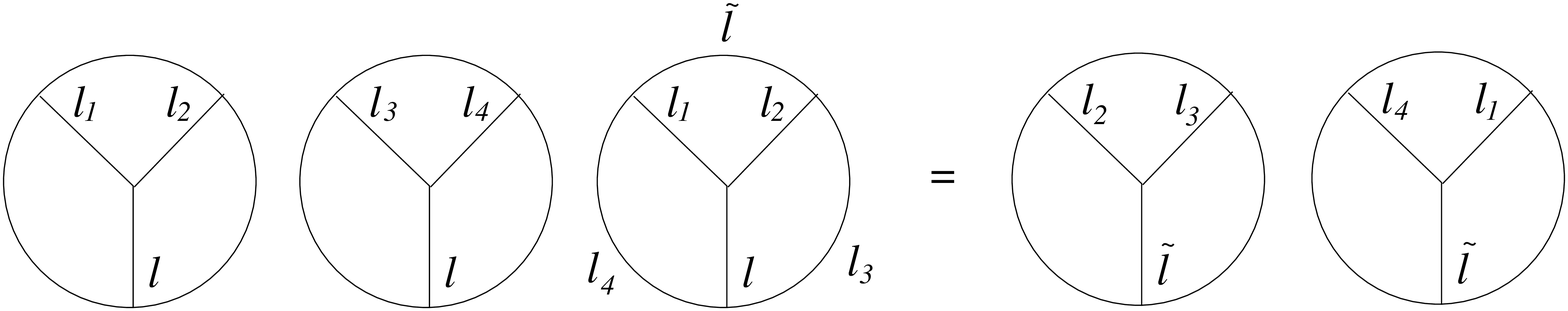,height=0.7in}}
\label{pentagon}
\end{equation}
Note that although one has to sum over all $l$ in this
formula, only the terms with $l\leq N$ survive.
We are also going to use the following recoupling identity:
\begin{equation}
\lower0.5in\hbox{\epsfig{figure=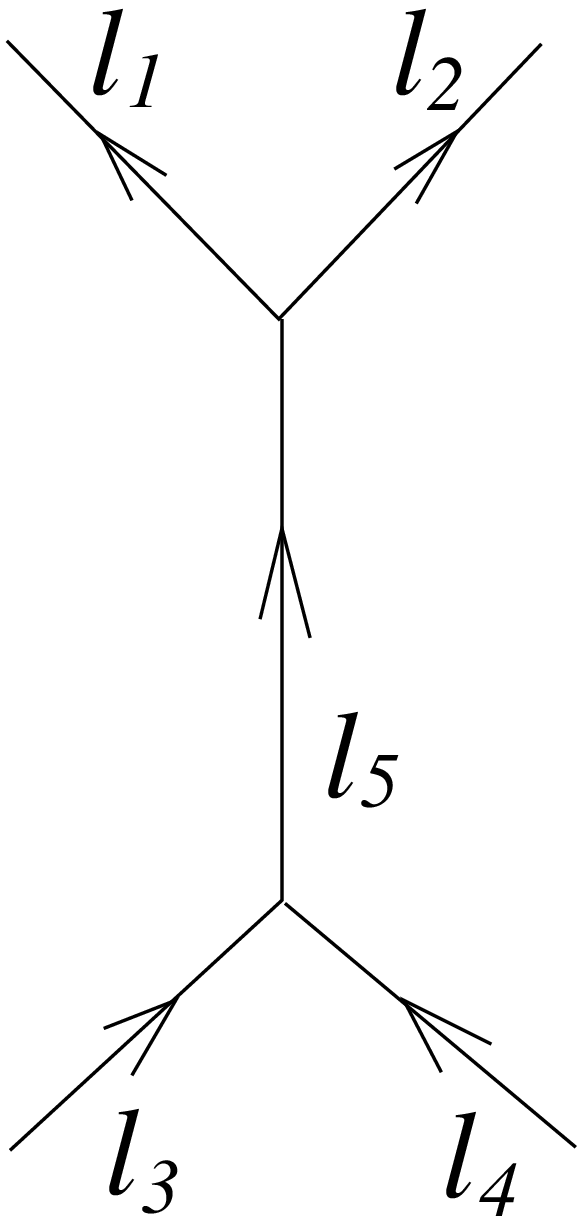,height=1in}} =
\sum_{l_6} {\rm dim}_{l_6} \,\,
\lower0.3in\hbox{\epsfig{figure=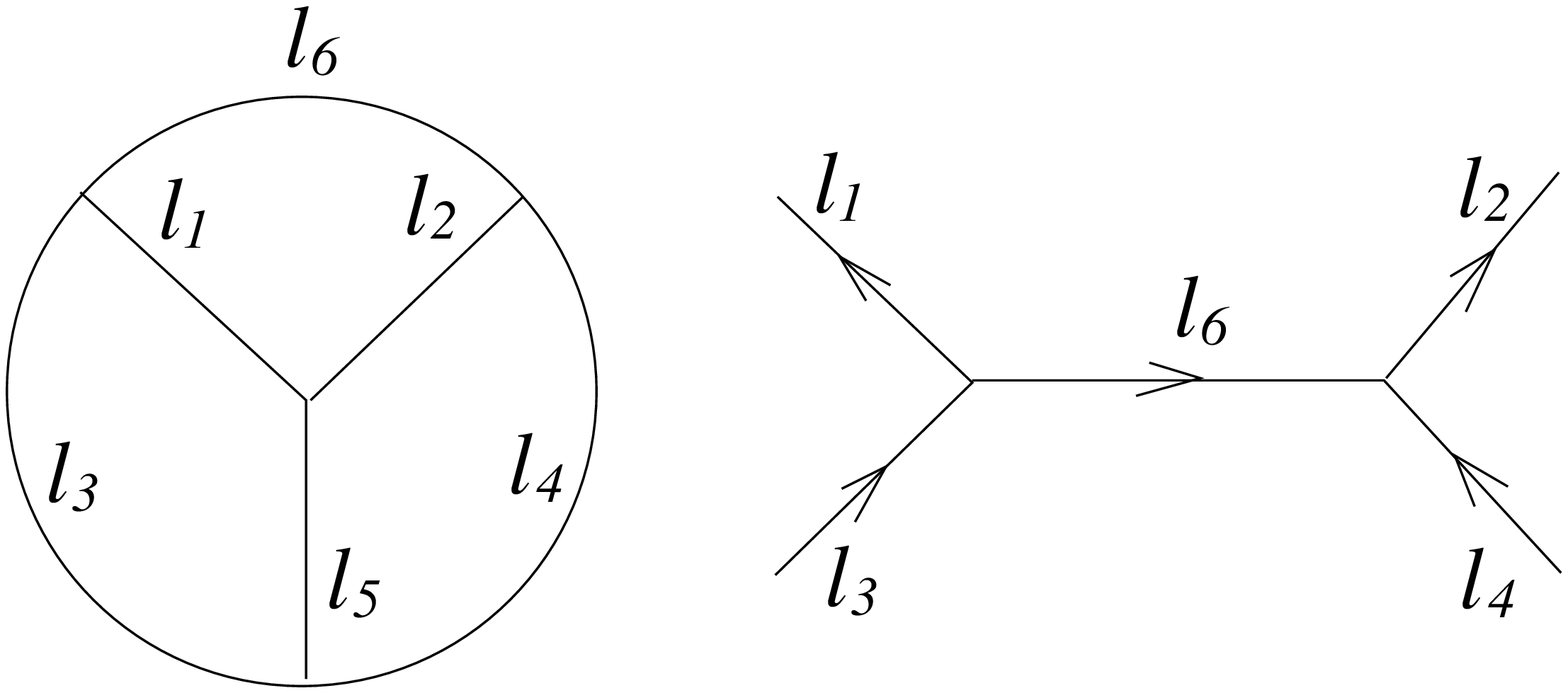,height=0.7in}}
\label{rec}
\end{equation}
Here the sum is taken over all $l_6$.
The proof of the associativity is then as follows:
\begin{eqnarray} \nonumber
\left( \t^{l_1} \star_N \t^{l_2} \right)
\star_N \t^{l_3} = \sum_{l_3'} {\rm dim}_{l_3'} \,
\t^{l_3'} \star_N \t^{l_3}
\lower0.3in\hbox{\epsfig{figure=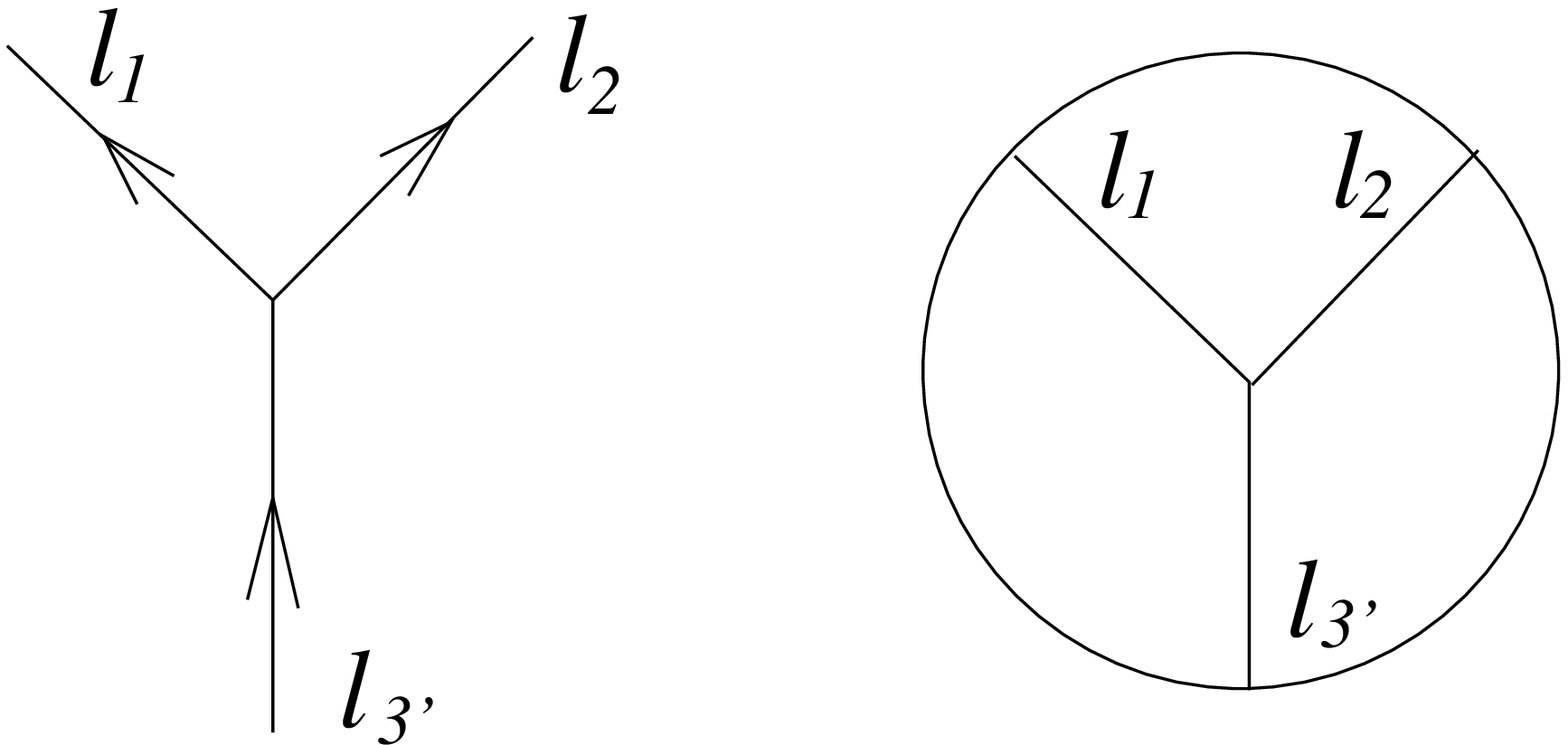,height=0.7in}} = \\ \nonumber
\sum_{l_3'\,l_4} {\rm dim}_{l_3'} {\rm dim}_{l_4}\, \t^{l_4}
\lower0.5in\hbox{\epsfig{figure=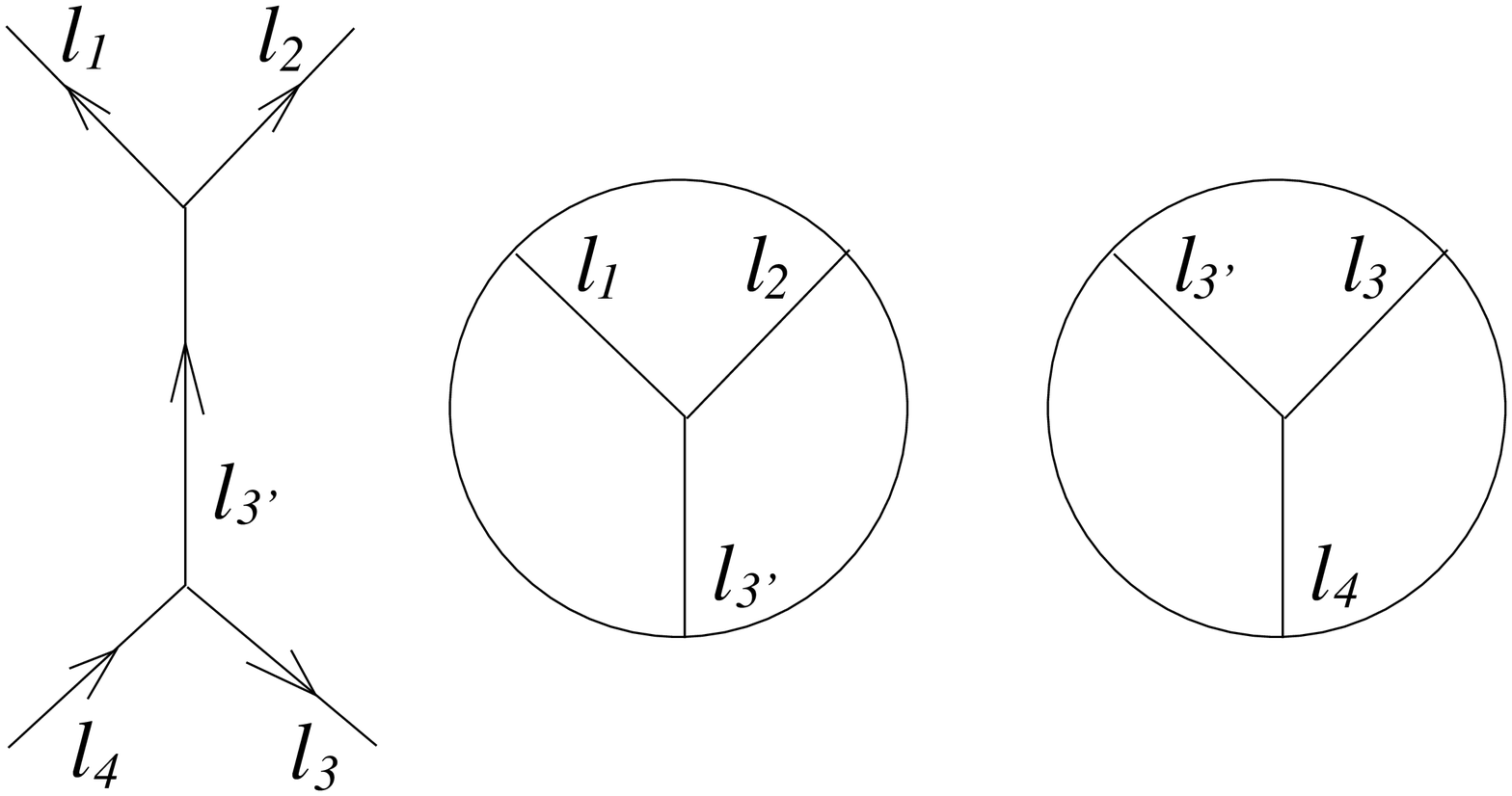,height=1in}} = \\ \label{ass-pr}
\sum_{l_3'\,l_3''\,l_4} {\rm dim}_{l_3'} {\rm dim}_{l_3''}
{\rm dim}_{l_4} \, \t^{l_4}
\lower0.3in\hbox{\epsfig{figure=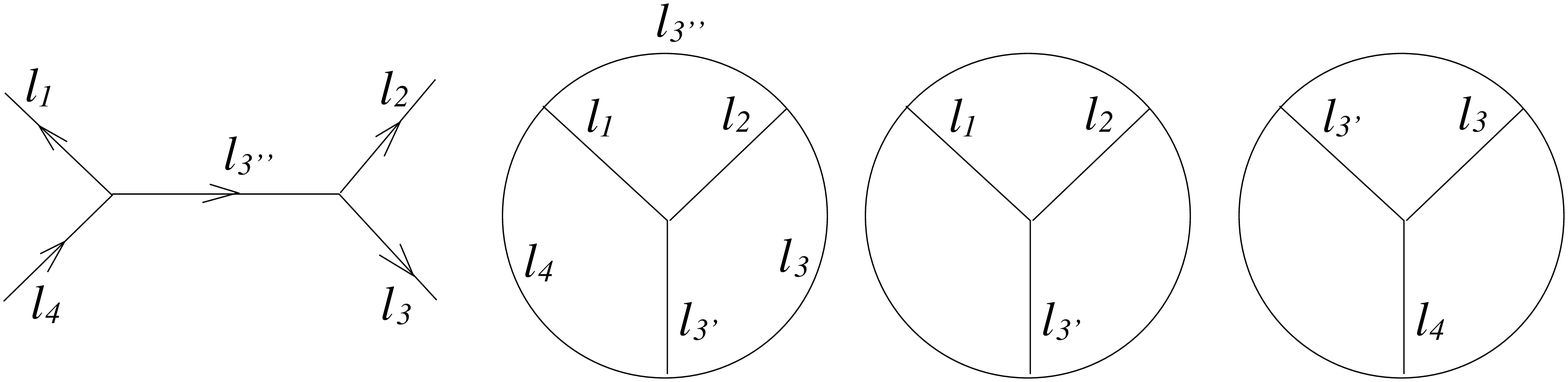,height=0.7in}} = \\ \nonumber
\sum_{l_3''\,l_4} {\rm dim}_{l_3''} {\rm dim}_{l_4} \, \t^{l_4}
\lower0.3in\hbox{\epsfig{figure=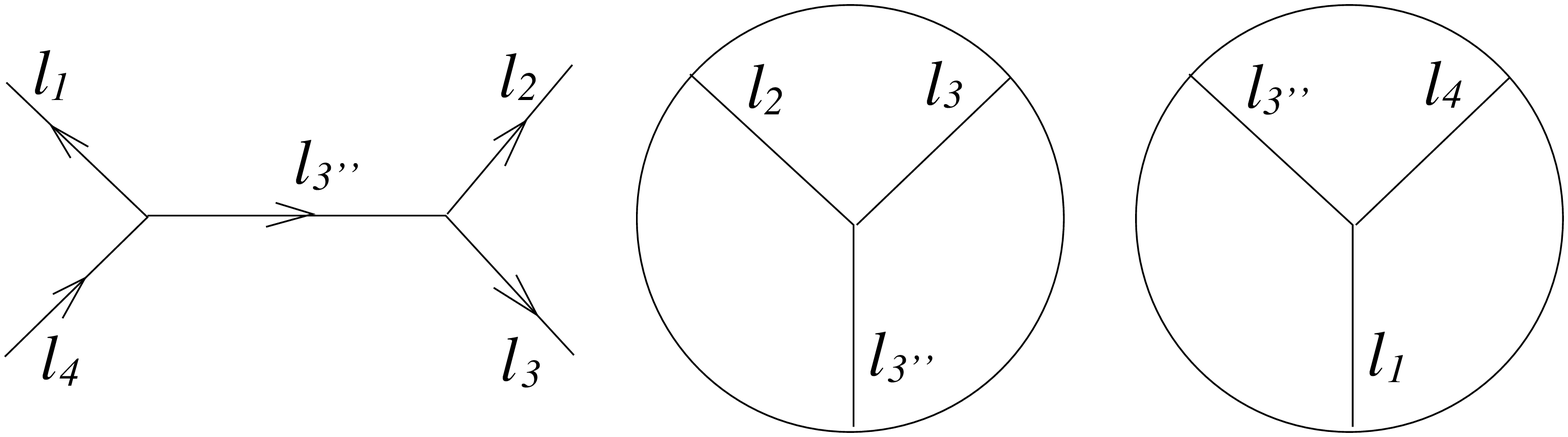,height=0.6in}} = 
\t^{l_1} \star_N \left( \t^{l_2} 
\star_N \t^{l_3} \right).
\end{eqnarray}
Here we used the recoupling identity (\ref{rec}) to get the
third line, and the pentagon identity (\ref{pentagon}) to
get the last line. Note that, although in the third line
the sum is taken over all $l_3''$, in the last line, after
we used the pentagon identity, only the terms with
$l_3''\leq N$ are non-zero. Thus, the associativity of the product
$\star_N$ is intimately related to the associativity in the
category of irreducible representations of ${\rm SU}(2)$, of
which the Biedenharn-Elliot identity is a manifestation.

Let us now study the asymptotic expansion of 
the $\star_N$-product more closely. To this end, let us rewrite the 
formula (\ref{app:6j}) for the $6j$-symbol in a form that
is convenient for taking the large $N$-limit. It is convenient to
introduce, for $N+a\geq 0$:\footnote{%
It is not hard to recognize this function as related to the one 
appearing in the infinite limit Euler definition of $\Gamma$ function.
Indeed, $\gamma(z;n)=\Gamma(z-1)/F(z-1,n+1)$, where $F(z,n)=n^z n!/
(z(z+1)\ldots(z+n))=n^z\int_0^1 (1-u)^n u^{z-1} du$, and 
$\lim_{n\to\infty} F(z,n)=\Gamma(z)$.}  
\begin{equation}
\gamma(a;N) = {(N+a)!\over N!(N+1)^a}.
\end{equation}
Since, for $a>0$,
\begin{equation}\label{gamma-zeros}
\gamma(a;N) = \prod_{k=1}^a (1+{k-1\over N+1}),
\end{equation}
this function, is analytic in $1/(N+1)$, and, for positive $a$, is 
a polynomial of degree $a-1$. We have: 
\begin{equation}\label{app:g2}
\gamma(a;N) = 1+{1\over N+1} {a(a-1)\over 2}+\ldots+
\left({1\over N+1}\right)^{a-1} (a-1)!
\end{equation}
One can similarly rewrite the function $\gamma(-a;N), a>0$ as
one over a polynomial of degree $a$:
\begin{equation}\label{gamma-poles}
\gamma(a;N) = {1\over\prod_{k=1}^a \left(1-{k\over N+1}\right)}.
\end{equation}
Using this function, the $6j$-symbol can be rewritten as:
\begin{eqnarray}\nonumber
&{}&\lower0.2in\hbox{\epsfig{figure=lll-2.eps,height=0.5in}} = 
(-1)^{l_1} (N+1)^{-1/2} l_3! \left[
{(l_1+l_2-l_3)!\over (l_1-l_2+l_3)! (l_2-l_1+l_3)! (l_1+l_2+l_3+1)! } 
\right]^{1/2} \\ \label{6j-gamma}
&{}&\left[
{\gamma(-l_1;N)\gamma(l_1+1;N)\gamma(l_3+1;N)\over 
\gamma(-l_2;N)\gamma(-l_3;N)\gamma(l_2+1;N)}
\right]^{1/2} \\ \nonumber
&{}&\sum_k {(-1)^k\over k!} {(l_1+k)! (l_2+l_3-k)!\over (l_1-k)! 
(l_2-l_3+k)! (l_3-k)!}
{\gamma(-l_3+k;N)\over \gamma(k+1;N)}.
\end{eqnarray} 
The sum here is restricted to those $k$ for which the quantities inside the 
factorials are non-negative. Using the representations (\ref{gamma-zeros}),
(\ref{gamma-poles}) for the $\gamma$ function, it is not hard to deduce the analyticity
properties of the $6j$-symbol as a function of $1/(N+1)$. We will
need these properties below, when we discuss the relation between
$\star_N$ and the usual $\star$-product. Function $\gamma(a+1;N), a>0$, viewed
as a function of $1/(N+1)$ has simple zeros on the negative axes, with
the closest to origin zero located at $1/(N+1) = - 1/a$. Function
$\gamma(-a;N), a$ has simple poles on the positive axes, with the
closest to the origin pole located at $1/(N+1)=1/a$.  Then, the
expression under the square root in (\ref{6j-gamma}) can be
shown to have at most simple poles and zeros on the positive axes
and at most second order zeros on the negative axes.
The function inside the sum in (\ref{6j-gamma}) has simple poles
both on the positive and negative axes. This proves that the 
$6j$-symbol times $\sqrt{N+1}$ (\ref{6j-gamma}) is an analytic 
function of $1/(N+1)$ in the open disc of radius 
${\rm min}(1/l_1,1/l_2,1/l_3)$ around zero. Moreover, all singularities
are located on the positive and negative axes. This means that the
function $\Psi_{(N)}(l_1,l_2,l_3)$ introduced in (\ref{Psi})
can be analytically continued, as a function of $1/(N+1)$, to the
whole complex plane. Introducing $\hbar=2/(N+1)$ we shall denote the
analytic continuation of $\Psi_{(N)}(l_1,l_2,l_3)$ by 
$\Psi(l_1,l_2,l_3;\hbar)$. 

It is not hard to find first terms of the expansion of $6j$
in powers of $1/(N+1)$. We get:
\begin{eqnarray}\nonumber
\sqrt{N+1}\,\,\lower0.2in\hbox{\epsfig{figure=lll-2.eps,height=0.5in}} = 
\hat{C}^{l_1\,l_2\,l_3}_{0\,0\,0}\left(
1+ {1\over (N+1)}{1\over 2} \left(C_{l_1} - C_{l_2} + C_{l_3}
\right)\right) \\ \label{6j-as-1}
 - {(-1)^{l_1}\over (N+1)} (l_3+1)!  \left[
{(l_1+l_2-l_3)!\over (l_1-l_2+l_3)! (l_2-l_1+l_3)! (l_1+l_2+l_3+1)! } 
\right]^{1/2} \\ \nonumber
 \sum_k {(-1)^k\over k!} k
{(l_1+k)! (l_2+l_3-k)!\over (l_1-k)! 
(l_2-l_3+k)! (l_3-k)!} + O(1/N^2),
\end{eqnarray}
where we have introduced a notation $C_l = l(l+1)$ and we have used 
the expression (\ref{app:3j}) for $C^{l_3 l_1 l_2}_{0\,0\,0}$.
This proves that the zeroth order term is given by the
usual commutative product.

Let us introduce a notation for the prefactor in front of the
sum:
\begin{eqnarray}
&{}& \rho({l_3 , l_1, l_2}) = (-1)^{l_1} l_3! \left[
{(l_1+l_2-l_3)!\over (l_1-l_2+l_3)! (l_2-l_1+l_3)! (l_1+l_2+l_3+1)! } 
\right]^{1/2}.
\end{eqnarray}
Then it can be noted that the sum in (\ref{6j-as-1}) is related
to a certain Clebsch-Gordan coefficient. Namely,
\begin{eqnarray}
&{}& \hat{C}^{l_3 l_1 l_2}_{1\;1\;0} ={1\over l_{3}} 
\left( {C_{3} \over C_{1}}\right)^{{1/2}}
\rho({l_3 , l_1, l_2})
\sum_k {(-1)^k\over k!} {(l_1+k)! (l_2+l_3-k)!\over (l_1-k)! 
(l_2-l_3+k)!} {k \over (l_3-k)!}.
\end{eqnarray}
Therefore, the first order term in the expansion, that is, the
coefficient of $1/(N+1)$ term in (\ref{6j-as-1}) can be written as:
\begin{equation}\label{expr-1}
    {1\over 2} (C_{1}- C_{2} +C_{3}) \hat{C}^{l_3 l_1 l_2}_{0\;0\;0}
    - (C_{1} C_{3})^{1/2} \hat{C}^{l_3 l_1 l_2}_{1\;1\;0}.
\end{equation}
It is now possible to prove the following two identities on 
Clebsch-Gordan coefficients, 
\begin{eqnarray}
(C_{1}- C_{2} +C_{3}) \hat{C}^{l_3 l_1 l_2}_{0\;0\;0} &=& 
(C_{1} C_{3})^{1/2} (\hat{C}^{l_3 l_1 l_2}_{1\;1\;0} + \hat{C}^{l_3 l_1 
l_2}_{-1\;-1\;0} ), \\ \nonumber
\hat{C}^{l_3 l_1 l_2}_{1\;1\;0}&{=}& (-1)^{l_{1}+l_{2}+l_{3}} 
\hat{C}^{l_3 l_1 l_2}_{-1\;-1\;0}.
\end{eqnarray}
The first equality is just the expression of the intertwining 
property of the Clebsch-Gordan coefficient, and the second is a 
symmetry relation.

These two identities imply that (\ref{expr-1}) is 
equal to zero if $l_{1}+l_{2}+l_{3}$ is even. It is also clear that
for odd $l_{1}+l_{2}+l_{3}$ it equals to 
\begin{equation}
     - (C_{1} C_{3})^{1/2} \hat{C}^{l_3 l_1 l_2}_{1\;1\;0}.
 \end{equation}
However, as is shown in the Appendix, this quantity is just
$P(l_1,l_2,l_3)$, which is the coefficient that appears in the
decomposition of the Poisson bracket of two spherical harmonics.
This proves that the first order term is given by the
Poisson bracket. 

We can summarize all of the above results
in a form of the following theorem.
\begin{theorem} 
There exists a function $\Psi(l_1,l_2,l_3;\hbar)$, given by the 
analytic continuation of the function $\Psi_{(N)}(l_1,l_2,l_3)$ introduced 
in (\ref{Psi}), which is analytic in $\hbar$ in the open disc of 
radius ${\rm min}(1/l_1,1/l_2,1/l_3)$ around zero, and  
\begin{equation}
\Psi (l_1,l_2,l_3;\hbar) = 1 + {\hbar} \Psi_1(l_1,l_2,l_3) +
\left( \hbar \right)^2 \Psi_2(l_1,l_2,l_3) + \ldots
\end{equation}
For $N\geq l_{1},l_{2},l_{3}$,  
\begin{equation}
    \Psi(l_1,l_2,l_3;\hbar)\Big|_{\hbar=2/(N+1)} = \Psi_{(N)}(l_1,l_2,l_3).
\end{equation}
The product defined via:
\begin{equation}\label{star-hbar}
\t^{l_1}_{m_1} \star_\hbar \t^{l_2}_{m_2} = 
\sum_{l_3=0}^N {\cal P}^{(l_3)} \left(
\t^{l_1}_{m_1} \cdot \t^{l_2}_{m_2} \right)
\Psi(l_1,l_2,l_3;\hbar)
\end{equation}
is an associative product. Moreover,
\begin{eqnarray}\nonumber
\phi\star_\hbar \psi - \psi\star_\hbar \phi = 
\hbar \{\phi,\psi \} + o(\hbar), \\
{\cal P}_N[\phi]\star_N {\cal P}_N[\psi] = 
{\cal P}_N \left[ \phi{\star_\hbar}\psi\right]
\Big |_{\hbar=2/(N+1)}.
\label{star-2}
\end{eqnarray}    
\end{theorem}
The associativity follows from the analytic continuation
of Biedenharn-Elliot identity, see (\ref{pentagon}). What we did not prove
is that the non-commutative product (\ref{star-hbar}) is
indeed a $\star$-product, that is given by the 
expansion in terms of derivatives. This is, in principle,
possible with our techniques by considering the higher
terms in the expansion of the $6j$-symbol. Thus, modulo
this caveat, the $\star_\hbar$-product gives a deformation
quantization of Kirillov bracket, and must thus be
equivalent to the usual Kirillov $\star$-product. The
formula (\ref{star-2}) then establishes a relation between
the fuzzy sphere product and Kirillov product.

We conclude by observing that the expressions
(\ref{6j}), (\ref{expr}) for the $\star_N$-product 
can be used to define a non-commutative product on the
so-called $q$-deformed fuzzy sphere. While the usual fuzzy
sphere is defined as the structure covariant under the action
of the group ${\rm SU}(2)$, its $q$-deformed analog is covariant
under the action of the quantum group $U_q(su(2))$. For a definition
of the $q$-deformed fuzzy sphere see, e.g., \cite{GMS}. The
expression (\ref{6j}) can then be used to get the product on the
$q$-deformed sphere. To this end, one should replace all objects
appearing on the right hand side of (\ref{6j}) --the dimension, the 
$3j$ and $6j$-symbols-- by the corresponding $q$-deformed quantities. 
One gets a $q$-deformed product. This product is still associative,
for the proof (\ref{ass-pr}) of associativity based on the pentagon 
identity holds for the quantum group as well. This 
$q$-deformed product was discussed in \cite{Alekseev}, where the
usual fuzzy sphere version (\ref{6j}) is also mentioned.
It would be quite interesting to obtain an analog of our
asymptotic formula for the $q$-deformed product.

\bigskip
\noindent
{\large \bf Acknowledgments}

K. K. was supported by the NSF grant PHY95-07065, L. F. was supported 
by CNRS and an ACI-Blanche grant.

\appendix
\section{Some formulas}

In the main text we use the values of the integrals of the product of two
and three matrix elements. These are given by:
\begin{equation}\label{int-2}
\int_{G} dg \, \overline{\langle l_1, m_1 | T_g | l_1, m_1' \rangle}
\langle l_2, m_2 | T_g | l_2, m_2' \rangle = {\delta^{l_1 l_2} \delta_{m_1 m_2}
\delta_{m_1' m_2'} \over {\rm dim}_{l_1}} ,
\end{equation}
and
\begin{equation}\label{int-3}
\int_{G} dg \, \overline{\langle l_3, m_3 | T_g | l_3, m_3' \rangle}
\langle l_1, m_1 | T_g | l_1, m_1' \rangle
\langle l_2, m_2 | T_g | l_2, m_2' \rangle =
\hat{C}^{l_3 l_1 l_2}_{m_3' m_1' m_2'} 
\overline{\hat{C}^{l_3 l_1 l_2}_{m_3 m_1 m_2}}.
\end{equation}

Let us now give some explicit formulas for the $3j$ and $6j$-symbols. 
All these formulas are from \cite{VK}, with normalizations properly 
adjusted to match our conventions. We have:
\begin{eqnarray}\nonumber
&{}&\lower0.2in\hbox{\epsfig{figure=lll-2.eps,height=0.5in}} = 
(-1)^{l_1} l_3! \left[
{(l_1+l_2-l_3)!\over (l_1-l_2+l_3)! (l_2-l_1+l_3)! (l_1+l_2+l_3+1)! } 
\right]^{1/2} \\ \label{app:6j}
&{}&\left[
{(N-l_1)!(N+l_1+1)!(N+l_3+1)!\over (N+l_2+1)! (N-l_2)! (N-l_3)!}
\right]^{1/2} \\ \nonumber
&{}&\sum_k {(-1)^k\over k!} {(l_1+k)! (l_2+l_3-k)!\over (l_1-k)! (l_2-l_3+k)! (l_3-k)!}
{(N-l_3+k)!\over (N+k+1)!}.
\end{eqnarray} 
Here the sum is taken over all $k$ such that the factorials are taken of 
non-negative integers. Using this formula it is not hard to get the
value (\ref{111}) of the $6j$-symbol for all spins being equal to 1.
In this case the sum in (\ref{app:6j}) is taken only over two values
$k=0,1$ and the calculation leading to (\ref{111}) is straightforward.

Let us also give an expression for the $3j$-symbol 
$C^{l_3 l_1 l_2}_{0\,0\,0}$. 
It can be obtained from the general expression for the
$3j$-symbol (in terms of a finite sum) given in \cite{VK}. 
Taking into account the difference in normalizations, we get:
\begin{equation}\label{app:3j}
\hat{C}^{l_3 l_1 l_2}_{0\,0\,0} = \rho({l_3 , l_1, l_2})
\sum_k {(-1)^k\over k!} {(l_1+k)! (l_2+l_3-k)!\over (l_1-k)! (l_2-l_3+k)! 
(l_3-k)!},
\end{equation}
where 
\begin{eqnarray}
&{}& \rho({l_3 , l_1, l_2}) = (-1)^{l_1} l_3! \left[
{(l_1+l_2-l_3)!\over (l_1-l_2+l_3)! (l_2-l_1+l_3)! (l_1+l_2+l_3+1)! } 
\right]^{1/2}.
\end{eqnarray}
We should also note the formula:
\begin{equation}\label{000}
\hat{C}^{l l_1 l_2}_{0\,0\,0} = {(-1)^{g-l} g! \Delta(l_1,l_2,l)\over
(g-l_1)!(g-l_2)!(g-l)!},
\end{equation}
where $l_1+l_2+l=2g, g\in\Z$, and $\Delta(l_1,l_2,l)$ is given by
\begin{equation}\label{Delta}
\Delta(l_1,l_2,l) = \left[
{(l_1+l_2-l)!(l_1-l_2+l)!(l_2-l_1+l)!\over(l_1+l_2+l+1)!}\right]^{1/2}.
\end{equation}

\section{Poisson bracket of spherical harmonics}
\label{app:poisson}

In this section we calculate the Poisson bracket of spherical harmonics.
The result we obtain here is compared in section \ref{sec:star} with
the first order term in the expansion of the $6j$-symbol in powers of $1/N$.
We were not able to find a result for this Poisson bracket in the
literature, so we sketch the calculation here.

We begin with some notations. Let $J_i$ be generators of the Lie
algebra of $\SU(2)$: $[J_i,J_j]=i\epsilon_{ijk} J_k$. These
generators can be realized as vector fields in $\R^3$. Denoting
by $x_i$ the usual Cartesian coordinates in $\R^3$, we get:
\begin{equation}
J_i = {1\over i} \epsilon_i^{\,\, jk} x_j \partial_k.
\end{equation}

To calculate the Poisson bracket, it is convenient to introduce a set of
complex coordinates in $\R^3$. We define
$z = (x_1+ix_2)/\sqrt{2}, x = x_3$ and 
$J_\pm = (J_1\pm i J_2)/\sqrt{2}, J = J_3$.
These new generators can be expressed in terms of the 
complex vector fields $\partial_z, \partial_\z, \partial_x$. We have:
\begin{equation}
J_+ = x \partial_\z - z \partial_x, \qquad
J_- = \z\partial_x + x \partial_z, \qquad
J = z\partial_z - \z\partial_\z.
\end{equation}
These vector fields satisfy:
\begin{equation}
[J_+,J_-] = J, \qquad
[J,J_{\pm}] = \pm J_{\pm},
\end{equation}
and 
\begin{equation}
\overline{J_+} = - J_-, \qquad \overline{J} = - J.
\end{equation}

The spherical harmonics $\t^l_m$ are given by:
\begin{equation}
\t^l_m = \alpha^l_m J_-^{l-m} v,
\end{equation}
where $v$ is the highest weight vector $v=z^l$. To calculate the
normalization factor $\alpha^l_m$ let us consider the norm $||J_-^{l-m} v||^2$.
Using 
\begin{equation}
[J_+,J_-^n]v = {n\over 2} (2l - n +1) J_-^{n-1} v,
\end{equation}
we get
\begin{equation}
||J_-^{l-m} v||^2 = (-1)^{l-m} \langle v| J_+^{l-m} J_-^{l-m}| v \rangle=
(-1)^{l-m} {1\over 2^{l-m}} {(l-m)!\over (l+m)!} (2l)! ||v||^2.
\end{equation}
An explicit calculation, using the normalized measure on $S^3$, gives:
\begin{equation}
||v||^2 = ||z^l||^2 = \int \sin^{2l}\theta = 
{1\over (2l+1)} {l! \over (2l-1)!!}.
\end{equation}
Combining these two facts, and taking into account the normalization
of $\t^l_m$ (\ref{norm}), we see that $\alpha^l_m$ is, up to a phase,
given by:
\begin{equation}
{1\over l!} \left( {(l+m)!\over 2^m (l-m)!} \right)^{1/2}.
\end{equation}
We choose the phase factor in such a way that $\t^l_m$ coincide with
the ones given by (\ref{theta}). This gives:
\begin{equation}
\alpha^l_m = {(-1)^{l-m} \over l!} 
\left( {(l+m)!\over 2^m (l-m)!} \right)^{1/2}.
\end{equation}

It is now straightforward to work out the action of vector fields
$\partial_z, \partial_\z, \partial_x$ on $\t^l_m$. We get:
\begin{eqnarray}\nonumber
\partial_z\t^l_m =  \sqrt{{1\over 2}(l+m)(l+m-1)} \t^{l-1}_{m-1}, \\
\label{vf}
\partial_\z\t^l_m = - \sqrt{{1\over 2}(l-m)(l-m-1)} \t^{l-1}_{m+1}, \\
\nonumber
\partial_x\t^l_m = \sqrt{(l+m)(l-m)} \t^{l-1}_m.
\end{eqnarray}
We will also need the action of $\SU(2)$ generators on $\t^l_m$:
\begin{eqnarray}\nonumber
J_+ \t^l_m &=& - \sqrt{{1\over 2}(l-m)(l+m+1)} \t^l_{m+1}, \\
\label{gen}
J_- \t^l_m &=& - \sqrt{{1\over 2}(l+m)(l-m+1)} \t^l_{m-1}, \\
\nonumber
J \t^l_m &=& m \t^l_m.
\end{eqnarray}

The Poisson bracket is given by:
\begin{equation}
\{f,g\} = {i\over\sin\theta}\left({\partial f\over\partial\theta}
{\partial g\over \partial\varphi} -{\partial g\over\partial\theta}
{\partial f\over \partial\varphi}\right).
\end{equation}
It is normalized so that $\{x_1,x_2\} = ix_3$. It can be expressed
in terms of vector fields and generators as:
\begin{equation}
\{f,g\} = \partial_z f (J_+ g) + \partial_\z f (J_- g) + 
\partial_x f (J g).
\end{equation}

Using the expressions (\ref{vf}), (\ref{gen}) it is now straightforward
to compute the Poisson bracket of two spherical harmonics. We have:
\begin{eqnarray}\nonumber
\{\t^{l_1}_{m_1},\t^{l_2}_{m_2}\} = - {1\over 2}\Big(
\sqrt{(l_1+m_1)(l_1+m_1-1)(l_2-m_2)(l_2+m_2+1)}
\t^{l_1-1}_{m_1-1}\t^{l_2}_{m_2+1} \\
- \sqrt{(l_1-m_1)(l_1-m_1-1)(l_2+m_2)(l_2-m_2+1)}
\t^{l_1-1}_{m_1+1}\t^{l_2}_{m_2-1} \\ \nonumber
- 2\sqrt{(l_1+m_1)(l_1-m_1)} \, m_2 \t^{l_1-1}_{m_1}\t^{l_2}_{m_2}\Big).
\end{eqnarray}
Thus, using (\ref{integral}) we have:
\begin{eqnarray}\nonumber
&{}&\int \{\t^{l_1}_{m_1},\t^{l_2}_{m_2}\} \overline{\t^{l}_{m}} =
- {1\over 2} 
\overline{\hat{C}^{l\, l_1-1\, l_2}_{0\, 0\,\,\,\,\,\, 0}} \\ \nonumber
&{}&\Big(
\sqrt{(l_1+m_1)(l_1+m_1-1)(l_2-m_2)(l_2+m_2+1)} 
\hat{C}^{l\,\, l_1-1\,\,\,l_2}_{m\, m_1-1\, m_2+1} \\ 
&{}&- \sqrt{(l_1-m_1)(l_1-m_1-1)(l_2+m_2)(l_2-m_2+1)}
\hat{C}^{l\,\, l_1-1\,\,\,  l_2}_{m\, m_1+1\, m_2-1} \\ \nonumber
&{}&- 2\sqrt{(l_1+m_1)(l_1-m_1)} \, m_2
\hat{C}^{l\,\,  l_1-1\,\, l_2}_{m\, m_1\, m_2} \Big).
\end{eqnarray}
The requirement of gauge invariance implies that the quantity in
brackets is proportional to the Clebsch-Gordan coefficient
$\hat{C}^{l\,l_1\,l_2}_{m\,m_1\,m_2}$, with the proportionality
coefficient depending only on $l_1, l_2, l$. In other words, we must have:
\begin{eqnarray}\label{bracket}
&{}&\int \{\t^{l_1}_{m_1},\t^{l_2}_{m_2}\} \overline{\t^{l}_{m}} =
P(l_1,l_2,l) \hat{C}^{l\,l_1\,l_2}_{m\,m_1\,m_2}.
\end{eqnarray}
The coefficient $P(l_1,l_2,l)$ is proportional to 
$\hat{C}^{l\, l_1-1\, l_2}_{0\, 0\,\,\,\,\,\, 0}$ and thus is non-zero only
when $l_1+l_2+l=2g-1, g\in\Z$. Since 
\begin{equation}
\hat{C}^{l\,l_2\,l_1}_{m\,m_2\,m_1} = (-1)^{l-l_1-l_2}
\hat{C}^{l\,l_1\,l_2}_{m\,m_1\,m_2},
\end{equation}
for values of $l_1,l_2,l$ summing up to an odd integer the Clebsch-Gordan
coefficient changes sign under the exchange of $l_1$ with $l_2$. Because
the Poisson bracket must be anti-symmetric, the coefficient $P(l_1,l_2,l)$
must be symmetric under the exchange of $l_1$ with $l_2$. Let us now 
determine this coefficient. It can be determined, for example, 
by choosing $l=m$ and using:
\begin{equation}
\hat{C}^{l\,l_1\,l_2}_{l\,j\,l-j} = (-1)^{l_1-j}
{(l_1+l_2-l)!\over(l_1+l_2+l+1)! \Delta(l_1,l_2,l)}
\left[
{(l_1+j)!(l_2+l-j)!\over (l_1-j)!(l_2-l+j)!}\right]^{1/2},
\end{equation}
where
\begin{equation}
\Delta(l_1,l_2,l) = \left[
{(l_1+l_2-l)!(l_1-l_2+l)!(l_2-l_1+l)!\over(l_1+l_2+l+1)!}\right]^{1/2}.
\end{equation}
After some algebraic manipulations this gives:
\begin{eqnarray}\label{P}
P(l_1,l_2,l) = (-1)^{{l_1+l_2-l+1\over 2}}
(2g+2){g!\Delta(l_1,l_2,l)\over (g-l_1)! (g-l_2)! (g-l)!}. 
\end{eqnarray}
Here $2g+1=l_1+l_2+l$ and we have used the formula (\ref{000}) for
$\hat{C}^{l\, l_1-1\, l_2}_{0\, 0\,\,\,\,\,\, 0}$. Expression 
(\ref{bracket}) together with (\ref{P}) is our final result for the
Poisson bracket of two spherical harmonics. Let us also note
that $P(l_1,l_2,l)$ can be written as a certain Clebsch-Gordan
coefficient:
\begin{equation}
P(l_1,l_2,l) = - \sqrt{l_1(l_1+1) l(l+1)} \hat{C}^{l l_1 l_2}_{1\;1\;0}.
\end{equation}
This equality is analogous to formula the (\ref{000}).

\end{document}